\documentclass[twocolumn]{aastex631}

\usepackage{amsmath}
\usepackage{amsthm}
\usepackage[ruled, noend, linesnumbered]{algorithm2e}
\let\oldnl\nl
\newcommand{\nlnonumber}{\renewcommand{\nl}{\let\nl\oldnl}}
\theoremstyle{remark}
\newtheorem*{remark}{Remark}

\usepackage{enumitem}
\usepackage{booktabs}      
\usepackage{multirow}

\newcommand{\R}{\mathbb R}
\newcommand{\N}{\mathbb N} 
\newcommand{\argmin}{\operatorname{argmin}}

\journalinfo{}

\begin{document}

\title{ImageMM: Joint multi-frame image restoration and super-resolution}

\author[0000-0002-4980-4291]{Yashil Sukurdeep}
\affiliation{Department of Applied Mathematics and Statistics, Johns Hopkins University, Baltimore, MD 21218, USA}

\author[0000-0002-7034-4621]{Tam\'{a}s~Budav\'{a}ri}
\affiliation{Department of Applied Mathematics and Statistics, Johns Hopkins University, Baltimore, MD 21218, USA}
\affiliation{Department of Physics and Astronomy, Johns Hopkins University, Baltimore, MD 21218, USA}
\affiliation{Department of Computer Science, Johns Hopkins University, Baltimore, MD 21218, USA}

\author[0000-0001-5576-8189]{Andrew J. Connolly}
\affiliation{DIRAC Institute and the Department of Astronomy, University of Washington, Seattle, WA 98195, USA}

\author{Fausto Navarro}
\affiliation{Johns Hopkins University, Baltimore, MD 21218, USA}

\correspondingauthor{Yashil Sukurdeep}
\email{yashil.sukurdeep@jhu.edu}

\begin{abstract}
A key processing step in ground-based astronomy involves combining multiple noisy and blurry exposures to produce an image of the night sky with an improved signal-to-noise ratio. Typically, this is achieved via image coaddition, and can be undertaken such that the resulting night sky image has enhanced spatial resolution. Yet, this task remains a formidable challenge despite decades of advancements.
In this paper, we introduce \emph{ImageMM}: a new framework based on the \emph{majorization-minimization} (MM) algorithm for joint multi-frame astronomical image restoration and super-resolution.
\emph{ImageMM} uses multiple registered astronomical exposures to produce a nonparametric latent image of the night sky, prior to the atmosphere's impact on the observed exposures. Our framework also features a novel variational approach to compute refined point-spread functions of arbitrary resolution for the restoration and super-resolution procedure. 
Our algorithms, implemented in \texttt{TensorFlow}, leverage graphics processing unit acceleration to produce latent images in near real time, even when processing high-resolution exposures.
We tested \emph{ImageMM} on Hyper Suprime-Cam (HSC) exposures, which are a precursor of the upcoming imaging data from the Rubin Observatory. The results are encouraging: \emph{ImageMM} produces sharp latent images, in which spatial features of bright sources are revealed in unprecedented detail (e.g., showing the structure of spiral galaxies), and where faint sources that are usually indistinguishable from the noisy sky background also become discernible, thus pushing the detection limits. Moreover, aperture photometry performed on the HSC pipeline coadd and \emph{ImageMM}'s latent images yielded consistent source detection and flux measurements, thereby demonstrating \emph{ImageMM}'s suitability for cutting-edge photometric studies with state-of-the-art astronomical imaging data.
\end{abstract}

\keywords{Astronomy image processing --- Ground-based astronomy --- GPU computing --- Majorization-minimization algorithm}

\section{Introduction} 
\label{sec:intro}
The widespread use of large-format detectors in astronomical projects has led to rapid growth in data volume and complexity in the field of astronomy, making it one of the most data-intensive fields of study today. Of particular interest to us are modern surveys where ground-based telescopes capture repeated observations of significant portions of the sky. These include the Hyper Suprime-Cam (HSC) survey~\citep{aihara2018hyper}, as well as the upcoming Legacy Survey of Space and Time (LSST) from the Rubin Observatory~\citep{ivezic2019lsst}. These ground-based surveys produce vast quantities of wide-field, deep-sky imaging data from which one can extract expansive amounts of information about the cosmos. 

A key step in processing images captured by ground-based telescopes involves combining multiple noisy and blurry astronomical exposures into a sharp, high-fidelity image of the night sky, ideally with improved spatial resolution. We will refer to this resulting image as a \emph{restoration}, or \emph{reconstruction}, of the night sky. 

Unfortunately, the aforementioned restoration task is typically hindered by several factors, chief among them being the varying levels of blur from exposure to exposure caused by changes in the atmosphere, the airmass, and the parallactic angle of observation. Other obstacles include the low signal-to-noise ratio of the exposures, their high dynamic range, and spurious or missing pixel values due to instrument artifacts or occlusions in the telescope's field of view. Moreover, observed exposures in modern surveys can contain tens of millions of pixels. While this implies more measurements from which to extract information, the sheer high-dimensionality of the imaging data may also become a major impediment for the development of computationally efficient pipelines to process and analyze the images.

\subsection{Related work}
\label{ssec:related_work}
In what follows, we highlight several existing methods that have been proposed for the challenging task of producing restorations of the night sky using multiple noisy and blurry registered ground-based exposures of the same part of the sky, which will be denoted by \mbox{$y = \{y^{(1)}, \dots, y^{(n)}\}$}.

Perhaps the simplest technique employed in practice is \emph{lucky imaging}, which involves choosing the observations with the lowest levels of blur, and subsequently adding them up to obtain a reconstruction~\citep{tubbs2003lucky, law2006lucky, brandner2016lucky}. Typically, over 90\% of all exposures are discarded during the selection process~\citep{law2006lucky}, which may be impractical in regimes where a limited quantity of data is available.

Image \emph{coaddition} is also widely utilized, whereby restorations are computed as a pixel-by-pixel weighted average of multiple input exposures~\citep{lucy1992co, fischer1994optimal, annis2014sloan, jiang2014sloan, zackay2017coaad}. Coaddition suppresses noise and certain outliers, yielding restored images (known as \emph{coadds}) with higher signal-to-noise ratios. However, since some of the exposures may have large blurs, the resulting coadds tend to lose sharpness, especially in comparison to restorations produced using lucky imaging.

Furthermore, a plethora of so-called \emph{deconvolution} techniques have been used for multi-frame astronomical image reconstruction. As part of these approaches, each noisy and blurry exposure $y^{(t)}$ is modeled as the convolution of the true, background-subtracted, noise-free latent image of the sky $x$, with a point-spread function (PSF) $f^{(t)}$, plus an additive noise term $\eta^{(t)}$ (see Section~\ref{sec:model_imaging_data} for details). In this setting, restorations are produced by estimating $x$ (the unknown latent image of the sky behind the atmosphere). The process of solving for this latent image when the PSFs \mbox{$f=\{f^{(1)}, \dots, f^{(n)}\}$} are known is referred to as \textit{deconvolution}. In contrast, estimating both the latent image and the PSFs when the latter are also unknown is called \textit{blind deconvolution}. A wide array of approaches have been proposed for astronomical image deconvolution, such as Bayesian methods based on maximum likelihood and maximum a posteriori estimation, as well as Fourier and wavelet-based deconvolution procedures; see the survey by~\cite{starck2002deconvolution} for a comprehensive overview. Several multi-frame blind deconvolution approaches also rely on maximum likelihood estimation~\citep{schulz1993multiframe, zhulina2006multiframe, matson2009fast}.

In the context of the aforementioned \emph{maximum likelihood estimation} (MLE) approaches, we highlight that obtaining a restoration of the night sky amounts to estimating $x$ (the true latent image of the sky) by finding an image $\widehat{x}$, called a \emph{maximum likelihood estimate}, which is most likely to have generated the observed exposures $y$ under a given statistical model. To solve for $\widehat{x}$, one typically optimizes a \emph{log-likelihood function}, which is derived based on assumptions on the distribution of pixel values in the additive noise terms $\eta^{(t)}$ (see Section~\ref{ssec:mle_background} for details). For instance, a Poisson noise assumption leads to the well-known \emph{Richardson-Lucy algorithm} and its variants~\citep{richardson1972bayesian, lucy1974iterative, fish1995blind}, while a Gaussian noise assumption (with constant variance across pixels) leads to the so-called \emph{Image Space Reconstruction algorithm} and its variants~\citep{daube1986iterative,law1996blind}. Thus, different distributional assumptions on the noise terms lead to different (blind) deconvolution algorithms, implying that MLE-based techniques provide a flexible, data-driven statistical framework for obtaining restorations of the night sky and estimating blurs.

Yet, MLE approaches typically fail to produce physically meaningful reconstructions, in particular when the log-likelihood optimization procedure is unconstrained~\citep{schulz1993multiframe}. Most prominently, unwanted artifacts often appear in the restored images, such as ringing caused by Gibbs oscillations~\citep{starck2002deconvolution}. 

In response, several methods attempt to constrain MLE-based (blind) deconvolution procedures via the addition of penalty terms when optimizing the log-likelihood function, giving rise to the so-called \emph{penalized MLE} techniques~\citep{schulz1993multiframe}. Other approaches involve the addition of regularizers on the maximum likelihood estimate via handcrafted priors on the distribution of its pixel values, leading to the so-called \emph{maximum a posteriori (MAP)} estimation techniques~\citep{starck2002deconvolution}. Penalized MLE and MAP methods constrain the set of feasible choices for the reconstruction $\widehat{x}$, which tends to yield physically meaningful restorations. However, one usually needs to employ algorithms based on (stochastic) gradient descent to optimize the penalized or regularized log-likelihood functions, resulting in procedures whose convergence properties and computational costs deteriorate as the number of pixels in the input exposures increases. In particular, such methods are impractical for processing high-resolution exposures, such as those produced by the LSST and HSC surveys.

More recently, \emph{streaming methods} for multi-frame (blind) deconvolution have been introduced~\citep{harmeling2009online, harmeling2010multiframe, hirsch2011online, lee2017streaming, lee2017robust}. Such frameworks have been used to jointly perform (blind) deconvolution and \textit{super-resolution}, which refers to the process of improving the spatial resolution of the latent image of the night sky. One of the major benefits of super-resolution is that it enables photometric and statistical tests to be performed on restored images at a higher resolution than that of the exposures. This can, for example, facilitate the detection of low-brightness objects at subpixel scales, and it can yield more precise measurements of sizes and distances between sources~\citep{lee2017robust}.

Similar to MLE approaches, the goal of streaming methods is to find the latent images (and possibly PSFs) that optimize a given log-likelihood function. The crucial difference between MLE and streaming frameworks lies in the approach for optimizing the log-likelihood. While MLE techniques \textit{directly} seek to optimize the log-likelihood function, usually via (stochastic) gradient descent, streaming methods operate by performing descent on an \textit{auxiliary function} of the (negative) log-likelihood via the \emph{expectation-maximization (EM)} algorithm, or its generalized version: the \emph{majorization-minimization (MM)} algorithm (see Section~\ref{ssec:mm_background} for additional details). For a wide range of log-likelihood functions, approaches that leverage the EM or MM algorithm give rise to an iterative \textit{multiplicative update} procedure for estimating the unknown latent image (and the PSFs in the case of blind deconvolution). This is particularly beneficial for astronomical image reconstruction, as the computational cost of the multiplicative updates scales favorably with respect to the number of pixels in the input exposures. Moreover, as part of EM or MM-based approaches, one can easily enforce desired constraints to obtain physically meaningful restorations, such as non-negative pixel values in the restored image of the sky. 

Yet, these methods still struggle in the context of modern surveys, largely due to the sheer size of the imaging data (e.g., when exposures contain tens of millions of pixels), and also due to the low signal-to-noise ratio of the exposures (e.g., when exposures contain large noise-dominated regions). Moreover, restored images produced via streaming methods depend on the arbitrary order in which input exposures are processed, which is a major drawback.

\subsection{Contributions}
\label{ssec:contributions}
Inspired by the aforementioned approaches, we present \emph{ImageMM}: a new framework based on the MM algorithm for joint multi-frame astronomical image restoration and super-resolution. With \emph{ImageMM}, we produce high-fidelity nonparametric latent images of the night sky, prior to the impact of the atmosphere on the observed exposures.

In particular, \emph{ImageMM} features a novel variational approach for obtaining refined PSF estimates of arbitrary resolution for use in the image restoration and super-resolution procedure. These refined PSFs can facilitate several photometric tasks, such as the detection of low-brightness objects at subpixel scales, thereby pushing the detection limits. 

Moreover, we have implemented \emph{ImageMM} in \texttt{TensorFlow}, which allows users to seamlessly leverage graphics processing unit (GPU) acceleration during computations. As a result, restored images can be obtained in near real time, even when processing high-resolution exposures containing tens of millions of pixels. 

To illustrate the capabilities and performance of the \emph{ImageMM} framework, we performed tests on HSC exposures, which are a precursor of upcoming imaging data from the Rubin Observatory. The results are very encouraging, and demonstrate that \emph{ImageMM} is suitable for use in the context of data processing pipelines for cutting-edge studies with real ground-based astronomical imaging data. We also tested \emph{ImageMM} on simulated data, which provided further validation for the method's capabilities.

\section{Modeling astronomical exposures}
\label{sec:model_imaging_data}
Let us begin by describing the data and setup for \emph{ImageMM}. We are given a set of coregistered ground-based exposures of the same part of the sky, denoted by \mbox{$y=\{y^{(1)}, \dots, y^{(n)}\}$}. For each \mbox{$t=1,\dots,n$}, the image $y^{(t)}\!\in\!\R^d$ is a $d$-dimensional column vector representing a noisy, blurry observation taken at time \mbox{$t$}, and we denote its pixel values as $y^{(t)}_i$ for \mbox{$i = 1,\dots, d$}. For notational simplicity, the ensuing mathematical presentation will be formulated for images represented as one-dimensional arrays (column vectors). However, all models, derivations and algorithms in this paper can easily be reformulated, and have been implemented in our software solution, for two-dimensional image arrays.

\begin{figure*}[htbp]
    \begin{center}
        \includegraphics[trim=2 9 2 20, clip, width=.9\textwidth]{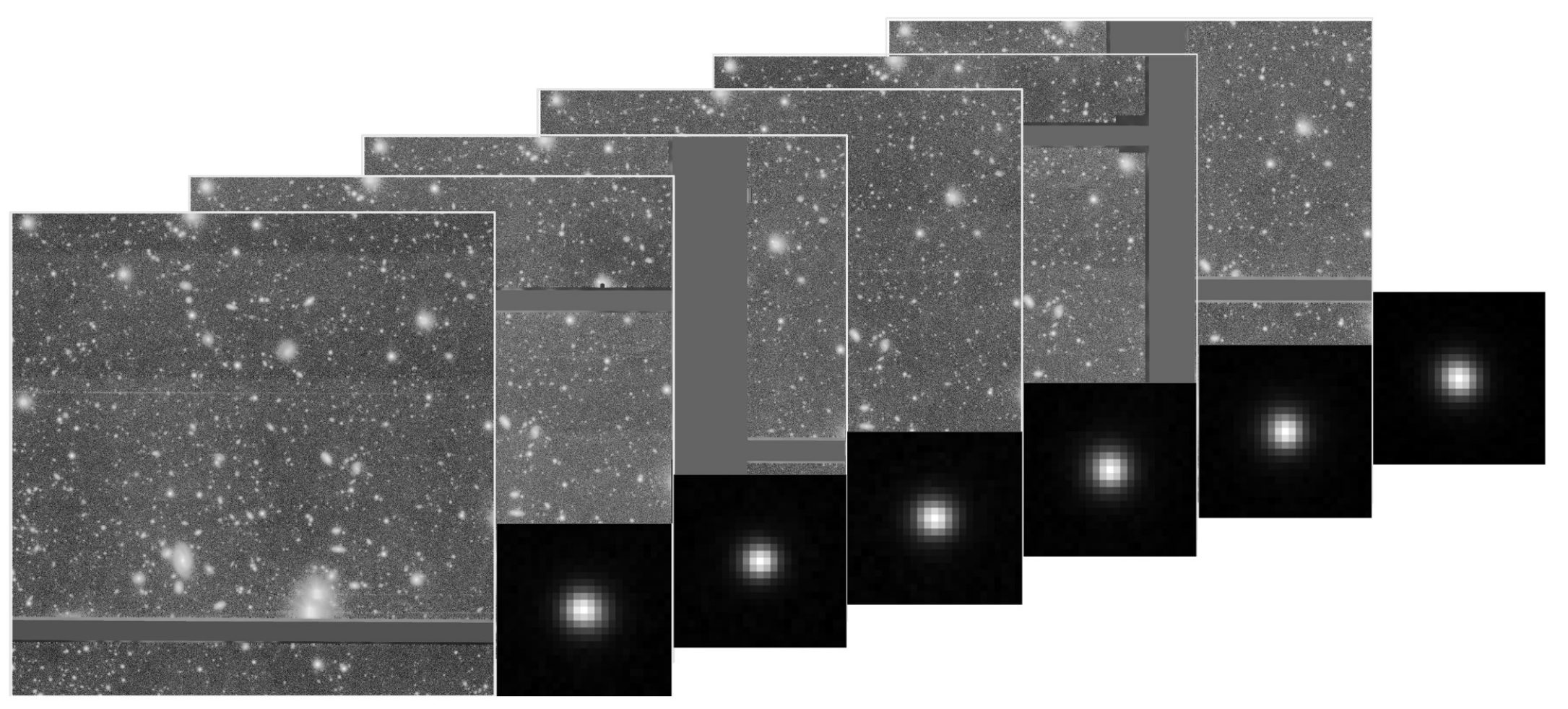}
    \end{center}
    \caption{\textbf{HSC $i$-band imaging data}, comprising a set of $n\!=\!33$ exposures \mbox{$y=\{y^{(1)}, \dots, y^{(n)}\}$}, each of size \mbox{$4200\!\times\!4200$} pixels (\mbox{$d = 4200^2$}), of which six are displayed above. Their corresponding PSFs \mbox{$f\!=\!\{f^{(1)}, \dots, f^{(n)}\}$}, each of size \mbox{$25 \!\times\! 25$} pixels (\mbox{$d' = 25^2$}), are displayed on the bottom right (not to scale). The large gray bands in the exposures represent chip gaps, which are examples of pixels for which \mbox{$m_i^{(t)} = 0$}.
    \label{fig:HSC_data}}
\end{figure*}

We note that the pixel values $y^{(t)}_i$ represent photon counts measured at each pixel in each exposure. We will assume that we are given corresponding standard deviations \mbox{$\sigma^{(t)}_i$}, and hence variances \mbox{$v_i^{(t)} \doteq (\sigma^{(t)}_i)^2$}, for these measurements. In addition, we are given corresponding point-spread functions (PSFs) \mbox{$f=\{f^{(1)}, \dots, f^{(n)}\}$} and \mbox{masks $m=\{m^{(1)}, \dots, m^{(n)}\}$} for each exposure. For each \mbox{$t = 1, \dots, n$}, the PSF \mbox{$f^{(t)}\!\in \R^{d'}$} is a $d'$-dimensional column vector (with \mbox{$d'\!< d$}) representing the convolution kernel (or blur) for image $y^{(t)}$. Typically, the PSFs are measured from stars in the exposures, which are selected from a catalog of sources. Meanwhile, the masks are binary-valued arrays encoding whether corresponding pixel values in the exposures are acceptable measurements. More precisely, for each \mbox{$t = 1, \dots, n$} and \mbox{$i = 1,\dots, d$}, the entries of the masks are defined as
\begin{equation*}
    m^{(t)}_i \doteq
    \begin{cases}
    1, \quad \text{if $y^{(t)}_i$ is an acceptable measurement,} \\
    0, \quad \text{otherwise.}
    \end{cases}
\end{equation*}

We note that the collection of exposures (with the variances of their pixel values), together with their associated PSFs and masks, correspond to a typical set of data products obtained from modern survey pipelines; see the HSC survey data in Figure~\ref{fig:HSC_data}, for instance.

With this data in hand, we model each observed exposure $y^{(t)}$ as the convolution between the true, background-subtracted, noise-free latent image of the night sky, denoted $x$, with the PSF $f^{(t)}$, plus an additive noise term $\eta^{(t)}$. The model for each observed pixel value in each exposure is thus
\begin{equation}
    \label{eq:model_exposures}
    y^{(t)}_{i} = ( f^{(t)} \!* x )_{i} + \eta^{(t)}_{i} .
\end{equation}
The pixel values in the noise terms $\eta^{(t)}_{i}$ are assumed to be \textit{independently} drawn samples from a probability distribution having mean zero and variance $v^{(t)}_{i}$. We emphasize that, in our model, the PSFs and noise terms can vary from exposure to exposure, while the underlying latent image of the sky is \emph{common to all exposures}.

\section{Mathematical background}
\label{sec:mathematical_background}
We now review key theoretical background on maximum likelihood estimation, which will allow us to subsequently introduce our majorization-minimization framework for astronomical image restoration.

\subsection{Maximum likelihood estimation}
\label{ssec:mle_background}
In this section, we focus on the scenario where our imaging data is modeled according to Equation~\eqref{eq:model_exposures}, and where the PSFs are \emph{known}, which we will assume to be the case for the remainder of this paper. In this setting, the task of producing a restored image of the night sky is a \textit{multi-frame deconvolution} problem, where the goal is to find the unknown, true latent image of the sky $x$. A natural way to estimate $x$ is via \textit{maximum likelihood estimation} (MLE), which involves finding an image $\widehat{x}$, called a \textit{maximum likelihood estimate}, that is most likely to have generated the observed exposures $y$ and the PSFs $f$ under our model. To find $\widehat{x}$, we minimize the joint negative log-likelihood of the pixel values of $x$ given the data (i.e., the exposures $y$ and PSFs $f$):
\begin{equation}
    \label{eq:deconvolution_mle_general}
    \widehat{x} = \underset{ x \in \mathcal{X} }{\operatorname{argmin}} ~{\mathcal{L}\left(x \mid y, f \right)}.
\end{equation}
In the formulation above, the specific functional form of the negative log-likelihood $\mathcal{L}$ typically depends on the distribution of the noise terms in Equation~\eqref{eq:model_exposures}; see Section~\ref{ssec:imagemm_restoration} and Section~\ref{ssec:imagemm_robust} for specific examples. Moreover, we note that the minimization in Equation~\eqref{eq:deconvolution_mle_general} takes place over the set of all images with non-negative pixel values, denoted by $\mathcal{X} \doteq \{ x \in \R_+^{d+d'-1} \}$. We impose this non-negativity constraint to obtain physically meaningful maximum likelihood estimates, in which pixels representing the sky have a value of zero, and where pixels representing sources (e.g., stars and galaxies) have strictly positive values. We also note that $\widehat{x}$ is padded (with \mbox{$d'\!-\!1$} extra pixels) in order to account for the influence of extra flux from sources outside the telescope's field of view when computing the reconstruction.

Typically, the constrained minimization in Equation~\eqref{eq:deconvolution_mle_general} is performed via (stochastic) gradient descent, which often converges to undesirable local minima, especially for imaging data with a large number of pixels. Thus, MLE-based multi-frame deconvolution methods often produce inadequate restorations~\citep{starck2002deconvolution}.

\subsection{Majorization-minimization}
\label{ssec:mm_background}
To address this issue, one can solve Equation~\eqref{eq:deconvolution_mle_general} by using the \textit{majorization-minimization} (MM) algorithm. In lieu of directly minimizing the negative log-likelihood function $\mathcal{L}$, the MM approach instead involves minimizing an \textit{auxiliary function} $\ell$ that \emph{majorizes} $\mathcal{L}$. Formally, this means that the auxiliary function possesses the following property for any pair of latent images $x, \Tilde{x} \in \R^{d+d'-1}$:
    \begin{equation}
        \label{eq:prop_auxliary_function}
       \ell(x \mid \Tilde{x}) \geq \ell(x \mid x ) = \mathcal{L}(x \mid y, f).
    \end{equation}
With such an auxiliary function in hand, one can indirectly minimize the negative log-likelihood by picking an initial guess $x_0$ for the maximum likelihood estimate, and constructing the following sequence of iterates until some convergence criterion is met:
    \begin{equation}
        \label{eq:mm_step}
        x_k = ~\underset{x \in \mathcal{X}}{\operatorname{argmin}} ~~\ell\left(x \mid x_{k-1} \right), \quad \text{ for } k \geq 1.
    \end{equation}
With an appropriate choice of initialization, the sequence of iterates $\{ x_k \}_{k \geq 0}$ converges to a maximum likelihood estimate for model~\eqref{eq:model_exposures}, i.e., it converges to a desired restoration of the night sky. This follows from the MM update rule~\eqref{eq:mm_step} and the properties of the auxiliary function~\eqref{eq:prop_auxliary_function}, which guarantee that the (negative) log-likelihood decreases at each successive iteration:
    \begin{eqnarray*}
        \mathcal{L}\left(x_{k-1} \mid y,f \right) & = & \ell\left(x_{k-1} \mid x_{k-1} \right) \\
        & \geq & \ell\left(x_k \mid x_{k-1} \right) \\
        & \geq & \ell\left(x_k \mid x_k \right) = \mathcal{L}\left(x_k \mid y, f\right) .
    \end{eqnarray*} 
The MM algorithm possesses several computational benefits that can be leveraged to great effect for astronomical image processing, as we shall describe next.

\section{The Image-MM framework}
\label{sec:imagemm}
Indeed, we now present \emph{ImageMM}, our novel framework based on the MM algorithm for multi-frame astronomical image restoration and super-resolution.

\subsection{Astronomical image restoration}
\label{ssec:imagemm_restoration}
Recall the model from Equation~\eqref{eq:model_exposures}, where for each \mbox{$t = 1, \dots, n$}, exposure $y^{(t)}$ is represented as the convolution of the true, background-subtracted, noise-free latent image of the sky, $x$, with the PSF $f^{(t)}$, to which the noise term $\eta^{(t)}$ is added. To fully define the model, we need to specify the distribution of the pixel values in the noise terms. To do so, we note that, while photon counts in the raw exposures follow a Poisson distribution, the large number of photons allows us to model pixel values in the noisy sky background, namely \mbox{$y^{(t)}_i - (f^{(t)} *  x)_i$} for \mbox{$i = 1, \dots, d$}, as \emph{independent}, \emph{mean-zero Gaussian random variables} whose variances are given by $v_i^{(t)} \doteq ( \sigma^{(t)}_{i} )^2$. Thus, we have that $\eta^{(t)}_{i} \sim \mathcal{N} \left(0, v^{(t)}_{i}\right)$. 

Under this modeling assumption, the joint negative log-likelihood of the pixel values of $x$, given $y$ and $f$, is obtained via the following sum-of-squares ($L_2$) loss function:
\begin{equation}
\label{eq:gaussian_likelihood_square_loss}
    \mathcal{L}(x\!\mid y, f ) = \frac{1}{2M} \sum_{t=1}^n \sum_{i=1}^d ~\! m_i^{(t)} \! \left( \frac{y_i^{(t)} \!- \left( F^{(t)} x \right)_i} {\sigma_i^{(t)}} \right)^2.
\end{equation}
The constant $M \doteq \sum_{t=1}^n \sum_{i=1}^d m_i^{(t)}$ is the total number of pixels with acceptable measurements across all exposures. Notice that, for notational purposes, we rewrite the convolution between PSF $f^{(t)}$ and latent image $x$ from Equation~\eqref{eq:model_exposures} as \mbox{$f^{(t)} \!* x = F^{(t)} x$}, where $F^{(t)}$ is the linear operator (i.e., matrix) corresponding to a convolution with kernel $f^{(t)}$; see~\citet{harmeling2010multiframe} for details.

Our approach to produce a restoration of the night sky thus consists of finding a maximum likelihood estimate for the Gaussian likelihood in Equation~\eqref{eq:gaussian_likelihood_square_loss}, or equivalently, computing a latent image $\widehat{x}$ that minimizes the $L_2$ loss in Equation~\eqref{eq:gaussian_likelihood_square_loss}, \emph{by using the MM algorithm}. To do so, we employ the following auxiliary function:
\begin{multline}
    \label{eq:auxiliary_l2_loss}
    \ell(x\!\mid\!\Tilde{x}) = \frac{1}{2M} \sum_{t=1}^n  \Bigg\{ y^{(t)}{}^{\top} W^{(t)}y^{(t)} - 2y^{(t)}{}^{\top} W^{(t)} F^{(t)}x \\ 
    +\Tilde{x}{}^{\top}F^{(t)}{}^{\top} W^{(t)} F^{(t)} \left( \frac{x \odot x}{\Tilde{x}} \right) \Bigg\},
\end{multline}
where $W^{(t)}$ is a diagonal $d \times d$ matrix whose $i^{th}$ diagonal entry is $W^{(t)}_{ii} \doteq m_i^{(t)} / v_i^{(t)}$ for all $i=1,\dots,d$, and where the multiplication ($\odot$) and division signs indicate pixel-by-pixel multiplication and division respectively. It can be verified that the auxiliary function above majorizes the negative log-likelihood from Equation~\eqref{eq:gaussian_likelihood_square_loss} as it satisfies the property in Equation~\eqref{eq:prop_auxliary_function}.

\begin{figure*}[htbp]
    \centering
    \includegraphics[trim={10mm 10mm 10mm 10mm},clip,width=.195\textwidth]{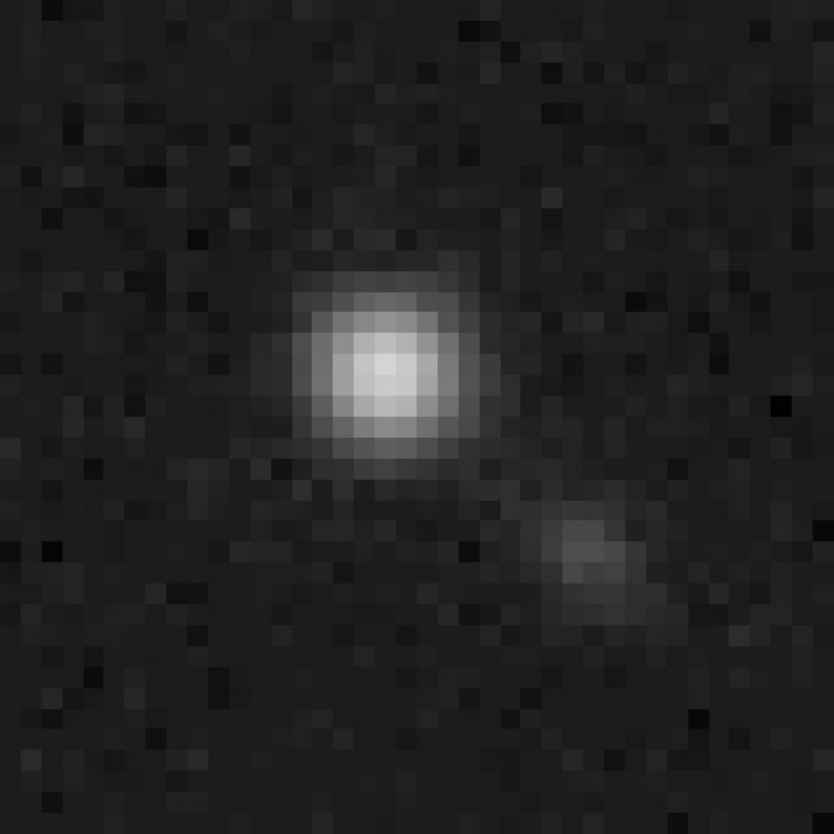}
    \includegraphics[trim={10mm 10mm 10mm 10mm},clip,width=.195\textwidth]{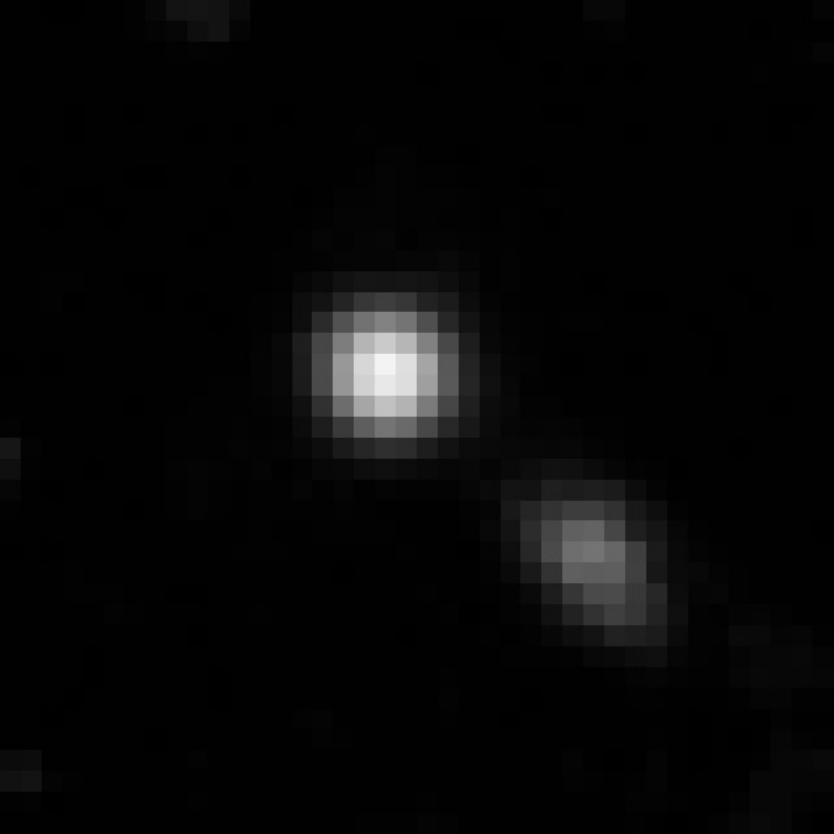}
    \includegraphics[trim={10mm 10mm 10mm 10mm},clip,width=.195\textwidth]{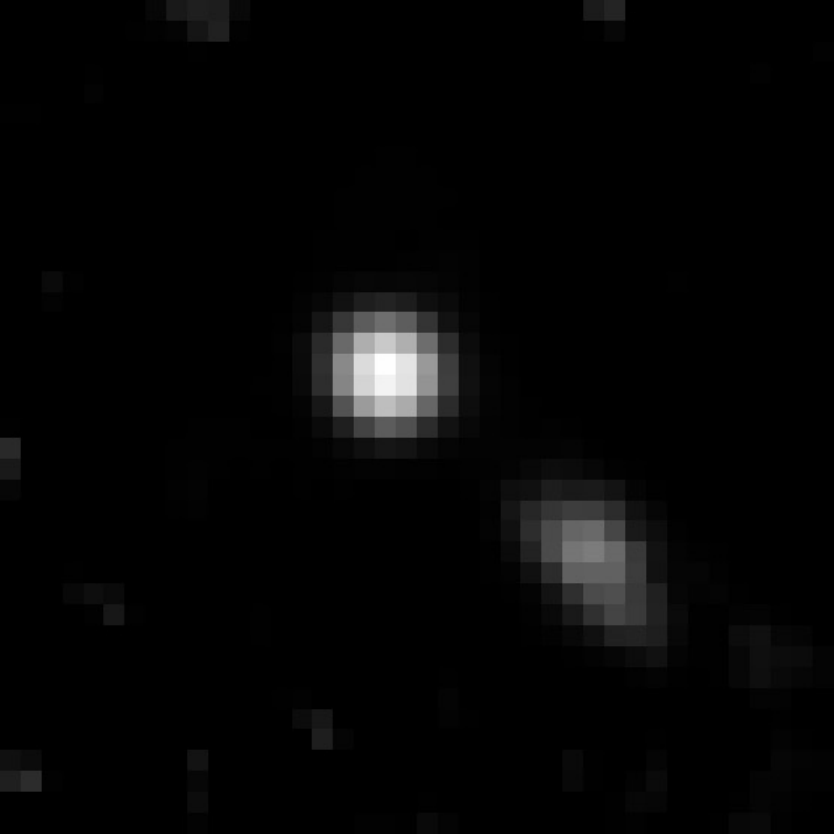}
    \includegraphics[trim={10mm 10mm 10mm 10mm},clip,width=.195\textwidth]{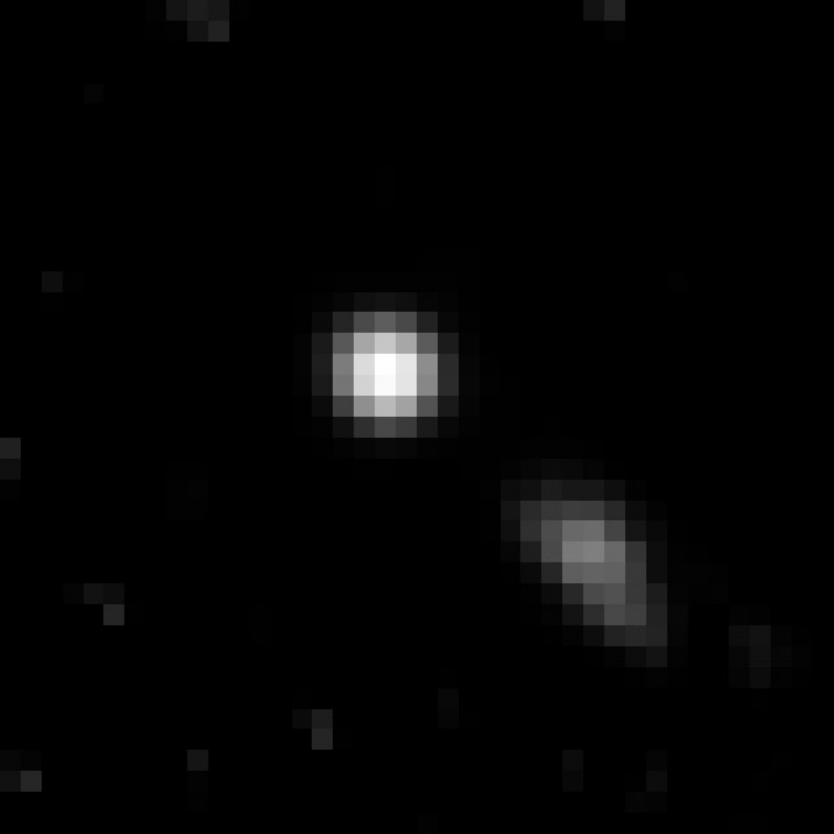}
    \includegraphics[trim={10mm 10mm 10mm 10mm},clip,width=.195\textwidth]{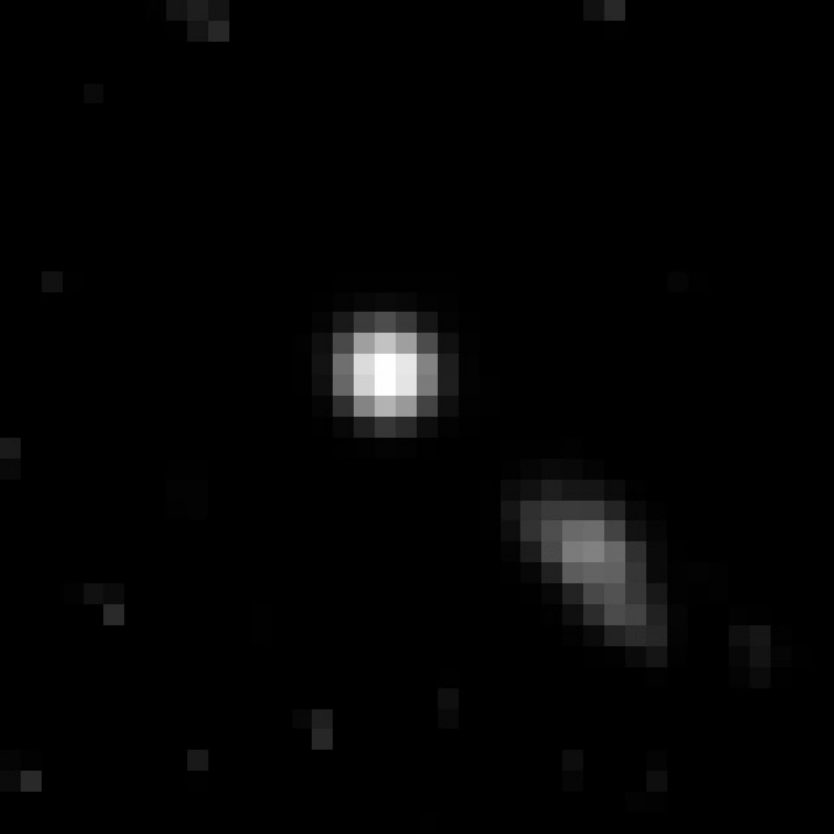}
    \caption{\textbf{Algorithm~\ref{alg:mm_multiframe_restoration} in action.} Our initial guess $x_0$ is a cutout of the median of the HSC exposures from Figure~\ref{fig:HSC_data} (\textit{left}). We iteratively apply the MM update formula (Equation~\eqref{eq:mm_update_l2_loss}), with update clipping (Equation~\eqref{eq:update_clipping}), to refine $x_0$. The updated guesses $x_k$ after \mbox{$k=5,10$,} and $15$ iterations are shown successively in the middle frames. The final restored image $\widehat{x}$ is obtained when the algorithm converges (\textit{right}). Notice how the algorithm progressively deblurs bright sources, while also removing noise in the sky background to reveal the potential presence of small, faint sources in $\widehat{x}$ that were not discernible in $x_0$.}
    \label{fig:x_hat_iterates}
\end{figure*}

We now have all the ingredients to present the core algorithm of the \emph{ImageMM} framework for multi-frame astronomical image restoration. Specifically, the algorithm entails choosing an initial guess $x_0$ for our estimate of the latent image of the night sky $\widehat{x}$, and iteratively updating this guess by applying the MM update rule in Equation~\eqref{eq:mm_step} with the auxiliary function from Equation~\eqref{eq:auxiliary_l2_loss}. More precisely, given our current guess for $\widehat{x}$ at the $k^{th}$ iteration, which we denote by $x_{k-1}$, we compute our updated guess $x_k$ by finding a stationary point of the auxiliary function, i.e., we find $x_k$ such that \mbox{$\nabla_x~\!\ell(x \mid x_{k-1}) \big|_{x=x_k} = 0$}. This leads to the following \emph{closed-form} expression for the updated guess (see Appendix~\ref{app:imagemm_mmupdatel2_formula} for a derivation):
\begin{equation}
    \label{eq:mm_update_l2_loss}
    x_{k} = x_{k-1} \odot u_k,
\end{equation}
where $u_k \in \R^{d + d' - 1}$ is the \emph{update image} defined as
\begin{equation}
    \label{eq:update_image_def}
    u_k \doteq \frac{ \sum_{t=1}^n {F^{(t)}}{}^{\top} W^{(t)} y^{(t)} } { \sum_{t=1}^n {F^{(t)}}{}^{\top} W^{(t)} F^{(t)} x_{k-1} } . 
\end{equation}

We summarize this procedure in Algorithm~\ref{alg:mm_multiframe_restoration}, and provide a visualization of the algorithm in action on a restoration task with HSC data in Figure~\ref{fig:x_hat_iterates}. Let us now highlight the key properties of Algorithm~\ref{alg:mm_multiframe_restoration}: 

\begin{algorithm}[htbp]
\SetKwInOut{Input}{Input}
\SetKwInOut{Output}{Output}
\SetKwFunction{FunctionMMRestoration}{ImageMMRestoration}

\caption{\emph{ImageMM} algorithm for multi-frame astronomical image restoration.}
\label{alg:mm_multiframe_restoration}

\DontPrintSemicolon

\Indm
\Input{Exposures, $y=\{y^{(1)}, \dots, y^{(n)}\}$. \\
PSFs, $f=\{f^{(1)}, \dots, f^{(n)}\}$. \\
Masks, $m=\{m^{(1)}, \dots, m^{(n)}\}$. \\
Variances $v = \{v_i^{(t)}\}$ for each pixel value $y_i^{(t)}$. \\
Initial guess for the latent image, $x_0$. \\
Maximum number of iterations, $K$. \\
Update clipping factor, $\kappa$.}

\BlankLine
\Output{Latent image of the night sky, $\widehat{x}$. \\
\noindent \hrulefill}

\nlnonumber
\FunctionMMRestoration{$y, f, m, v, x_0, K, \kappa$}:
\BlankLine

\Indp
Initialize $\widehat{x} \gets x_0$\;
\For{$t=1,\dots,n$}
    {$W^{(t)} \gets \operatorname{diag}\left(m_1^{(t)} / v_1^{(t)}, \dots,  m_d^{(t)} / v_d^{(t)}\right)$}
\While{$k \gets 1$ \KwTo $K$}
    {\Repeat{$u_k' \approx u_{k-1}'$}
    {$ u_k \gets \frac{ \sum_{t=1}^n {F^{(t)}}{}^{\top} W^{(t)} y^{(t)} } { \sum_{t=1}^n {F^{(t)}}{}^{\top} W^{(t)} F^{(t)} \widehat{x} }$\;
    \BlankLine
    $ u_k' \gets \max{\left\{ 1/ \kappa, ~\min{( \kappa, u_k)} \right\}}$\;
    \BlankLine
    $\widehat{x} \gets \widehat{x} \odot u_k' $\;
    \BlankLine}
    }
\Return{$\widehat{x}$}
\end{algorithm}

\medskip

\noindent \textbf{$\bullet$ Multiplicative update:} First, the update formula (Equation~\eqref{eq:mm_update_l2_loss}) only involves \emph{element-wise} multiplication of the pixel values of the current iterate $x_{k-1}$ with those of the update image $u_k$. At each iteration, the computational cost of this operation is thus \emph{linear} with respect to the number of pixels in the latent image, which allows us to obtain restorations with fast processing times.

Of course, one needs to compute the update image $u_k$ itself at each iteration, which requires convolutions and can thus be computationally expensive for images with a large number of pixels. Nevertheless, with GPU acceleration, these convolution operations can be performed rapidly even when processing high-resolution exposures containing several millions of pixels; see Section~\ref{ssec:implementation_performance} for further details about computation times with exposures of different sizes.

\medskip

\noindent \textbf{$\bullet$ Enforcing non-negativity in $\widehat x$:} Second, our MM procedure allows us to easily obtain a latent image $\widehat x$ with non-negative pixel values, which is desirable for physical interpretability as outlined in Section~\ref{ssec:mle_background}. Indeed, if the initial guess $x_0$ has \textit{strictly positive} pixel values and all the update images $u_k$ also have \textit{strictly positive} pixel values, then so will the restored image $\widehat x$, precisely due to the multiplicative update (Equation~\eqref{eq:mm_update_l2_loss}). 

This observation gives us several principled and practical initialization strategies for the procedure. For instance, one can choose $x_0$ to be a constant image where all pixels are set to a fixed positive value, or compute $x_0$ as the mean or median of the input exposures $y$, with appropriate padding and with replacement of its nonpositive pixel values if necessary. 

To maintain non-negativity in the update images, we follow the approach of~\citet{lee2017robust} and clip update images $u_k$ during the multiplicative update procedure. The \emph{clipped update images} are defined as follows:
    \begin{equation}
        \label{eq:update_clipping}
        u_k' \doteq \max{\left\{1 / \kappa, ~\min{(\kappa, u_k)}\right\}},
    \end{equation}
where \mbox{$\kappa > 1$} is the so-called \emph{update clipping factor}. The updates are more conservative when $\kappa$ is close to 1, and the clipping has virtually no impact when $\kappa$ is large. 

While our procedure is robust to the choice of initialization $x_0$ and update clipping factor $\kappa$, we have empirically observed that initializing with the mean or median of the input exposures and performing update clipping with \mbox{$\kappa = 2$} typically speeds up convergence to high-fidelity, physically meaningful latent images of the night sky, as depicted in Figure~\ref{fig:x_hat_iterates}.

\medskip

\noindent \textbf{$\bullet$ Convergence criterion:} Speaking of convergence, a natural criterion for determining when to terminate the procedure is to check when each entry of update image $u_k$ is roughly equal to $1$. Indeed, due to the multiplicative nature of the update formula (Equation~\eqref{eq:mm_update_l2_loss}), it would imply that the current iterate $x_{k-1}$ has converged. 

However, since we employ update clipping in practice, some of the entries of the clipped update images $u_k'$ might never approach a value of $1$. We thus determine convergence by instead checking whether $u_k' \approx u_{k-1}'$. In other words, the algorithm converges when \emph{the clipped update images themselves stop fluctuating}. We include a detailed discussion of the practical implementation of this aforementioned stopping criterion in Appendix~\ref{app:stopping_criteria_implementation}.

\medskip

\noindent \textbf{$\bullet$ Processing all frames simultaneously:} Moreover, a defining feature of our method is that we process all exposures $y\!=\!\{y^{(1)}, \dots, y^{(n)}\}$ \textit{simultaneously} when updating our estimate for the latent image $\widehat{x}$ using the MM update formula (Equation~\eqref{eq:mm_update_l2_loss}). 

This contrasts sharply with other existing multi-frame deconvolution methods that leverage the EM or MM algorithm, such as those of~\citet{harmeling2009online},~\citet{harmeling2010multiframe},~\citet{hirsch2011online},~\citet{lee2017robust} and~\citet{lee2017streaming}, in which the updates for $\widehat{x}$ are performed in a \emph{streaming} manner by processing one exposure at each iteration. While streaming can be desirable in settings where processing memory is limited, the resulting restoration of the night sky \emph{depends on the order in which input exposures are processed}. 

This is undesirable in regimes where one has access to a small number of exposures, or when exposures have very low signal-to-noise ratios, as is often the case with large-scale ground-based astronomical imaging data. Unhampered by this shortcoming, \emph{ImageMM} is thus well-adapted for producing high-fidelity restorations in the context of modern astronomy surveys.

\subsection{Super-resolution}
\label{ssec:super_resolution}
A natural extension of the \emph{ImageMM} framework is \textit{super-resolution}, where the goal is to produce a restoration of the night sky whose resolution is $r$ times that of the observed exposures, where $r \!\geq\! 1$ is called the \emph{super-resolution factor}.

More formally, let us consider the same setup as in Section~\ref{sec:model_imaging_data}, where we are given a set of coregistered exposures $y$ (and variances for their pixel values), together with corresponding PSFs $f$ and masks $m$. Recall that, for each \mbox{$t = 1,\dots,n$}, exposure \mbox{$y^{(t)} \!\in \R^d$} has a resolution of $d$ pixels (with the same being true for the mask $m^{(t)}$), while the resolution of PSF \mbox{$f^{(t)} \!\in \R^{d'}$} is $d'$ pixels, where \mbox{$d' \!< d$}. Consider a super-resolution factor $r \geq 1$, which we shall assume to be an integer for simplicity. 

With a view toward performing super-resolution, we consider a modified version of the model in Equation~\eqref{eq:model_exposures}, in which each observed exposure \mbox{$y^{(t)} \in \R^d$} is now modeled as the convolution of a \emph{super-resolved} latent image of the sky, \mbox{$x \in \R^{r(d + d') - 1}$}, with a \emph{super-resolved} PSF \mbox{$h^{(t)} \!\in \R^{rd'}$}, plus an additive noise term \mbox{$\eta^{(t)} \!\in \R^d$}. The model for each pixel value in each exposure is thus
\begin{equation}
    \label{eq:model_exposures_super_resolution}
    y^{(t)}_{i} = ( D (h^{(t)} \!* x ))_{i} + \eta^{(t)}_{i},
\end{equation}
where \mbox{$D: \R^{rd} \to \R^d$} is a \emph{downsampling operator} that acts on $rd$-dimensional column vectors by performing \emph{average pooling} with windows of size $r$. For now, we will also assume that \mbox{$\eta^{(t)}_{i} \sim \mathcal{N} \left(0, v^{(t)}_{i}\right)$} for \mbox{$i = 1,\dots,d$}, just as in Section~\ref{ssec:imagemm_restoration}. We emphasize that, in the model in Equation~\eqref{eq:model_exposures_super_resolution}, the latent image $x$ contains \mbox{$r(d+d')\!-\!1$} pixels, as opposed to the \mbox{$d+d'\!-\!1$} pixels in the model in Equation~\eqref{eq:model_exposures}. We therefore require PSFs $h^{(t)}$ with a (higher) resolution of $rd'$ pixels to model the blur in each exposure, as opposed to PSFs $f^{(t)}$ with $d'$ pixels for the model in Equation~\eqref{eq:model_exposures}.

Our approach for performing joint image restoration and super-resolution thus essentially entails solving for the super-resolved PSFs \mbox{$\{h^{(1)}, \dots, h^{(n)}\}$}, and subsequently feeding them as input to Algorithm~\ref{alg:mm_multiframe_restoration} in order to produce a super-resolved estimate $\widehat{x}$ for the latent image of the night sky. 

\begin{figure*}[htbp]
    \centering
    \includegraphics[width=.9\textwidth]{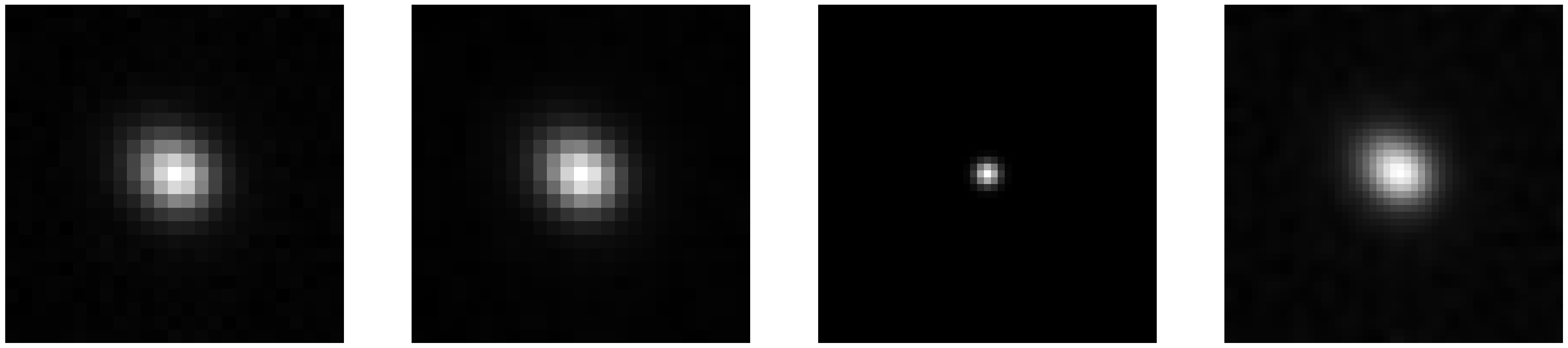}
    \caption{\textbf{Computing PSFs via Equation~\eqref{eq:psfs_super_resolution}.} \textit{Left to right:} (i) Target HSC PSF $f^{(t)}$, (ii) $\widehat{f}^{(t)} = D(h^{(t)} * g_{\sigma})$, (iii) Gaussian PSF $g_{\sigma}$, with $\sigma=1.1$, and (iv) super-resolved PSF $h^{(t)}$ with $r=2$. The average (squared) difference between pixel values of $f^{(t)}$ and $\widehat{f}^{(t)}$ is $3.94 \times 10^{-8}$, showing that our PSF-solving strategy produces PSFs satisfying Equation~\eqref{eq:h_psf_optimality} by finding global minimizers of Equation~\eqref{eq:psfs_super_resolution}.}
    \label{fig:psf_solver}
\end{figure*}

To this end, we introduce the following PSF-solving strategy, whereby for each \mbox{$t=1,\dots,n$}, we compute the PSF $h^{(t)}$ via the following variational problem:
\begin{equation}
    \label{eq:psfs_super_resolution}
    h^{(t)} = \underset{h \in \R^{rd'}}{\operatorname{argmin}}~\left\{ \frac{1}{d'}\sum_{i=1}^{d'} \left( f^{(t)}_i - \left( D \left( h * g_{\sigma} \right) \right)_i \right)^2 \right\},
\end{equation}
where \mbox{$g_{\sigma} \in \R^{rd' + rd' - 1}$} is a centered Gaussian PSF with standard deviation \mbox{$\sigma > 0$}, and where \mbox{$D: \R^{rd'} \to \R^{d'}$} is the downsampling operator introduced earlier. In practice, we solve this unconstrained optimization problem using the Adam algorithm~\citep{kingma2014adam}. By virtue of being a moderately sized problem (where typically $d'=O(10^2)$, as is the case for the HSC PSFs from Figure~\ref{fig:HSC_data} in which $d'=225$), the Adam optimizer consistently converges to global minimizers of Equation~\eqref{eq:psfs_super_resolution}, as illustrated in Figure~\ref{fig:psf_solver}. Additionally, we generate the Gaussian PSF $g_{\sigma}$ via Monte Carlo integration for enhanced numerical accuracy and stability when solving Equation~\eqref{eq:psfs_super_resolution}.

Moreover, notice that each PSF $h^{(t)}$ from Equation~\eqref{eq:psfs_super_resolution} satisfies
\begin{equation}
    \label{eq:h_psf_optimality}
    f^{(t)} = D ( h^{(t)} \!* g_{\sigma} ).
\end{equation}
In other words, our PSF-solving strategy produces super-resolved PSFs $h^{(t)}$ (of arbitrary resolution) that are equal to the target PSFs $f^{(t)}$ \emph{up to a convolution with $g_{\sigma}$} (after downsampling). Such PSFs are well-adapted for use in astronomical image processing. 

\begin{figure*}[htbp]
    \centering
    \includegraphics[width=0.29\textwidth]{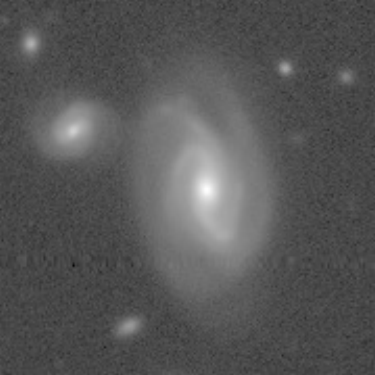}
    \includegraphics[width=0.29\textwidth]{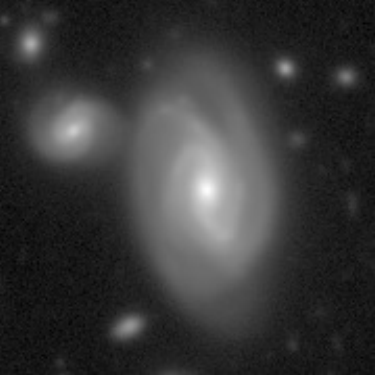}
    \includegraphics[width=0.29\textwidth]{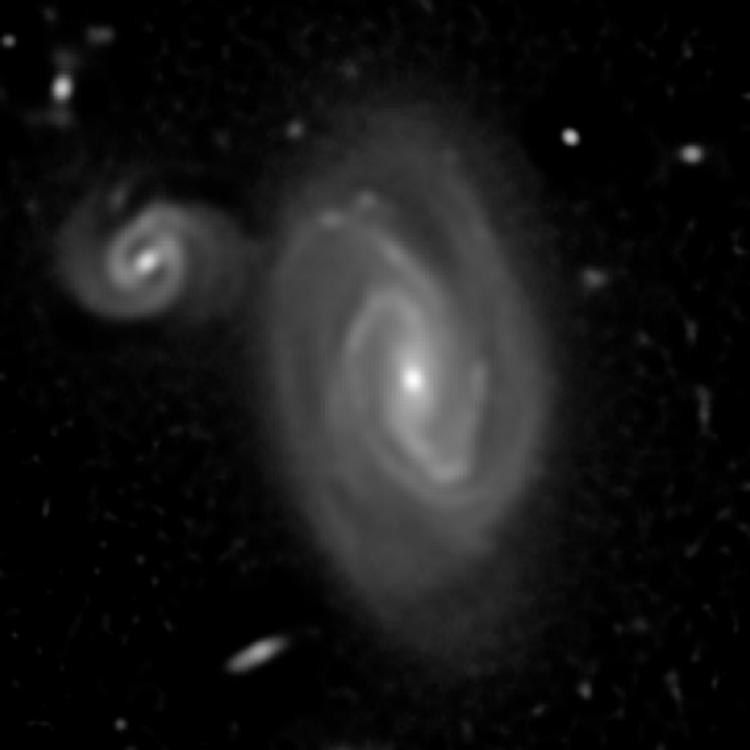}
        \caption{\textbf{Bright sources in restorations.} \textit{Left:} Cutout of an HSC \emph{exposure} containing several bright sources, in which the sky background is noisy. \textit{Middle:} \emph{Pipeline coadd} of the HSC exposures, in which sky-background noise is reduced, but where bright sources are still blurry. \textit{Right:} Our \emph{super-resolved latent image} $\widehat{x}$ computed with Algorithm~\ref{alg:mm_multiframe_restoration_superresolution}, which has twice the native resolution of the exposures ($r=2$), and was produced using $h$ PSFs computed via Equation~\eqref{eq:psfs_super_resolution} with $\sigma=1.1$. There is virtually no sky-background noise in the super-resolved restoration $\widehat{x}$, and bright sources appear significantly sharper than in the coadd, thereby revealing fine spatial features (e.g., the shape and structure of the spiral arms of the galaxies) in unprecedented detail.} 
        \label{fig:galaxy_restoration}
\end{figure*}

Indeed, the super-resolved PSFs enhance the efficacy of the deblurring process for bright sources, resulting in super-resolved latent images in which their spatial features are revealed in unprecedented detail, as illustrated by the structure of spiral galaxies shown in Figure~\ref{fig:galaxy_restoration}.

\begin{figure*}[htbp]
    \centering
    \includegraphics[width=.29\textwidth]{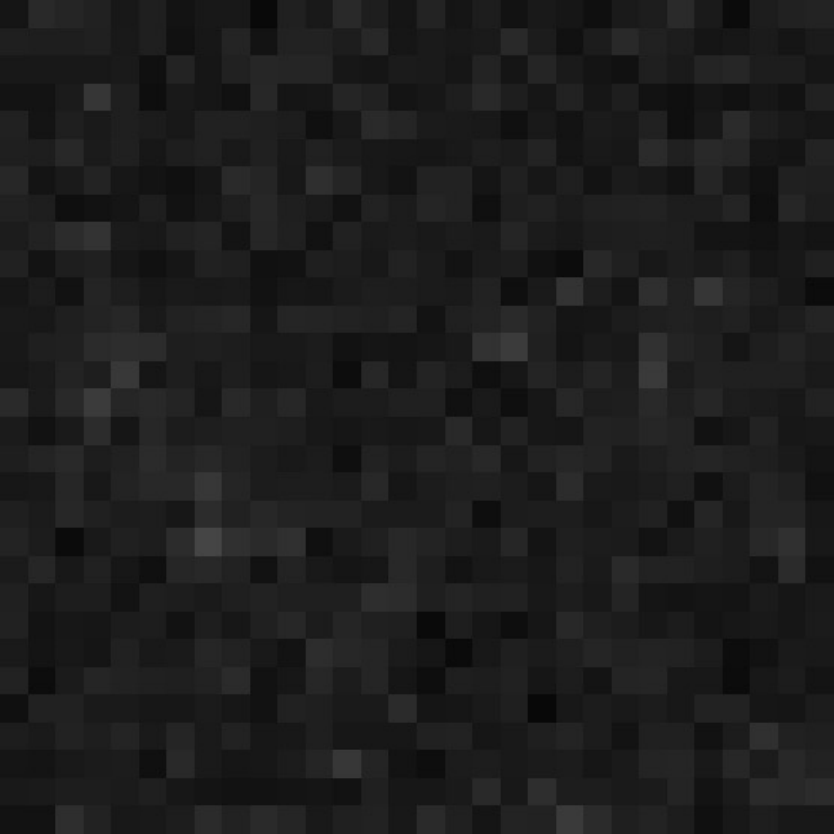}
    \includegraphics[width=.29\textwidth]{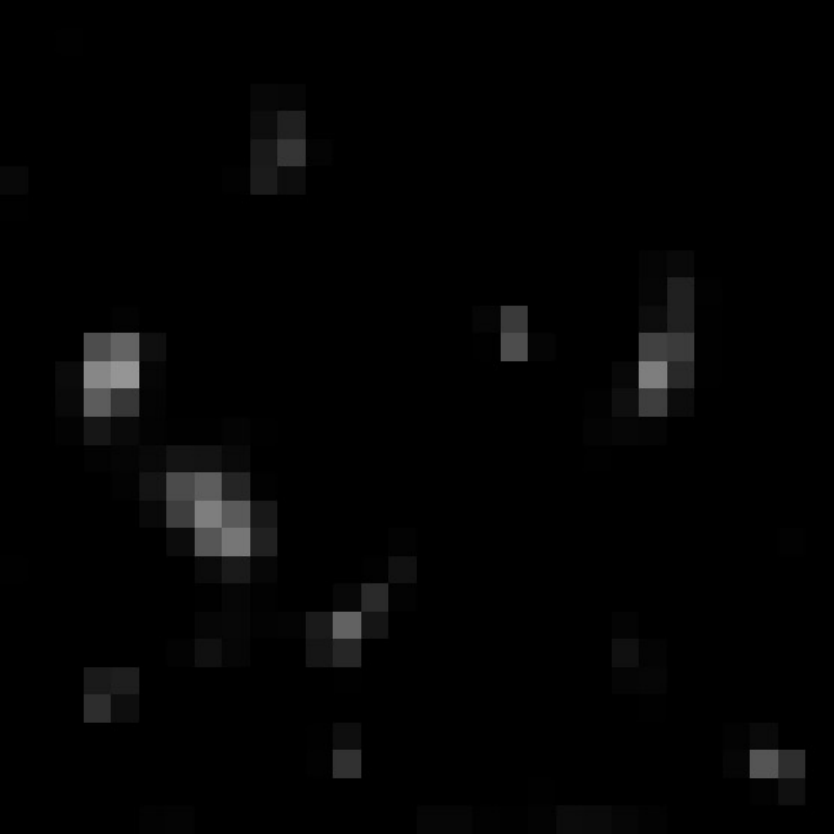}
    \includegraphics[width=.29\textwidth]{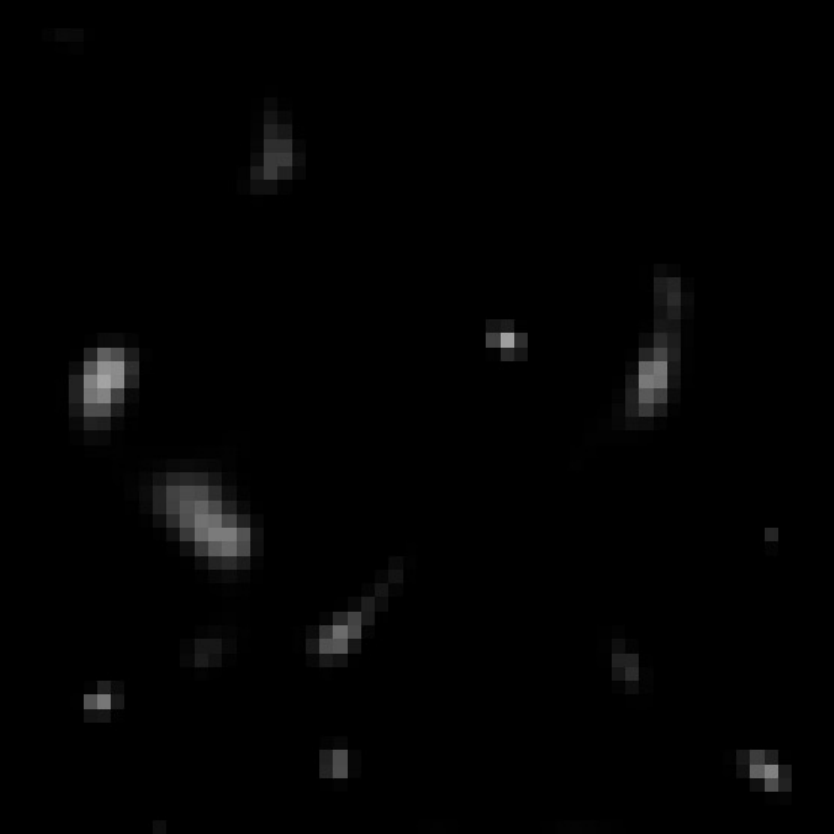}
    \caption{\textbf{Super-resolution facilitates the detection of faint sources.} \textit{Left:} Cutout of a faint region of the sky from the \emph{pipeline coadd} of the HSC exposures, in which faint, low-brightness sources are virtually indistinguishable from the noisy sky background. \textit{Middle:} \emph{Latent image} with \emph{native resolution} produced via Algorithm~\ref{alg:mm_multiframe_restoration_superresolution}, where we used $h$ PSFs computed via Equation~\eqref{eq:psfs_super_resolution} with $\sigma=1$ and $r=1$, i.e., \emph{no super-resolution}. Several faint sources now become clearly discernible due to the removal of sky-background noise in the latent image. \textit{Right:} Our \emph{super-resolved restoration} $\widehat{x}$ produced using Algorithm~\ref{alg:mm_multiframe_restoration_superresolution} with a super-resolution factor of $r\!=\!2$. With joint restoration \emph{and} super-resolution, yet more faint point sources become visible (at the subpixel scale) in the latent image. In particular, their shapes, sizes, and relative distances become more easily discernible, thus enhancing our detection capabilities.}
    \label{fig:lbo_restoration}
\end{figure*}

Moreover, the PSFs produced via Equation~\eqref{eq:psfs_super_resolution} are equally adept at deconvolving small, faint sources in darker regions of the sky. Indeed, by virtue of satisfying Equation~\eqref{eq:h_psf_optimality}, each PSF $h^{(t)}$ is sharper than its corresponding counterpart $f^{(t)}$, and thus represents a convolution kernel that induces a lower degree of blur in an image when compared to PSF $f^{(t)}$. Consequently, performing deconvolution with $h^{(t)}$ produces a latent image of the sky $\widehat{x}$ containing some extra level of blur compared to a deconvolution with $f^{(t)}$. This additional blur in $\widehat{x}$, which is determined by the Gaussian kernel $g_{\sigma}$ and can be controlled by the choice of the parameter $\sigma$, can for instance facilitate the detection of faint point sources at subpixel scales, in particular on super-resolved latent images of the night sky, as shown in Figure~\ref{fig:lbo_restoration}.

We provide a summary the \emph{ImageMM} framework for joint multi-frame astronomical image restoration and super-resolution in Algorithm~\ref{alg:mm_multiframe_restoration_superresolution}.

\begin{algorithm}[h!]
\SetKwInOut{Input}{Input}
\SetKwInOut{Output}{Output}
\SetKwFunction{FunctionMMRestorationPlus}{ImageMMRestorationPlus}

\caption{\emph{ImageMM} algorithm for joint multi-frame astronomical image restoration and super-resolution.}
\label{alg:mm_multiframe_restoration_superresolution}

\DontPrintSemicolon

\Indm
\BlankLine
\Input{Exposures, $y=\{y^{(1)}, \dots, y^{(n)}\}$. \\
PSFs, $f=\{f^{(1)}, \dots, f^{(n)}\}$. \\
Masks, $m=\{m^{(1)}, \dots, m^{(n)}\}$. \\
Variances $v = \{v_i^{(t)}\}$ for each pixel value $y_i^{(t)}$. \\
Super-resolution factor, $r$. \\
Standard deviation for Gaussian PSF, $\sigma$. \\
Initial guess for the latent image, $x_0$. \\
Maximum number of iterations, $K$. \\
Update clipping factor, $\kappa$.}
\BlankLine

\Output{Latent image of the night sky, $\widehat{x}$. \\
\noindent \hrulefill}

\nlnonumber
\BlankLine
\FunctionMMRestorationPlus{$y, f, m, v, r, \sigma, x_0, K, \kappa$}:
\BlankLine

\Indp
Initialize $\widehat{x} \gets x_0$\;
\For{$t=1,\dots,n$}
    {\BlankLine
    $W^{(t)} \gets \operatorname{diag}\left(m_1^{(t)} / v_1^{(t)}, \dots,  m_d^{(t)} / v_d^{(t)}\right)$
    \BlankLine}
{Generate Gaussian PSF $g_{\sigma} \in \R^{rd'+rd'-1}$ via Monte Carlo integration} \;
\For{$t=1,\dots,n$}
    {\BlankLine
    $h^{(t)} = \underset{h \in \R^{rd'}}{\operatorname{argmin}}~\frac{1}{d'}\sum_{i=1}^{d'} \left( f^{(t)}_i - \left( D \left( h * g_{\sigma} \right) \right)_i \right)^2 $
    \BlankLine}
\While{$k \gets 1$ \KwTo $K$}
    {\BlankLine
    \Repeat{$u_k' \approx u_{k-1}'$}
    {$ u_k \gets \frac{ \sum_{t=1}^n {H^{(t)}}{}^{\top}D^{\top} W^{(t)} y^{(t)} } { \sum_{t=1}^n {H^{(t)}}{}^{\top} D^{\top} W^{(t)} D H^{(t)} \widehat{x} }$\;
    \BlankLine
    $ u_k' \gets \max{\left\{ 1/ \kappa, ~\min{( \kappa, u_k)} \right\}}$\;
    \BlankLine
    $~\widehat{x} \gets \widehat{x} \odot u_k' $\;
    \BlankLine}
    }
\BlankLine
\Return{$\widehat{x}$}
\BlankLine
\end{algorithm}

\begin{remark} We point out the following:
\begin{itemize}
    \item In Algorithm~\ref{alg:mm_multiframe_restoration_superresolution}, we denote the linear operator corresponding to a convolution with PSF $h^{(t)}$ as $H^{(t)}$, and therefore, we have that $$H^{(t)}z = h^{(t)} * z$$ for all \mbox{$z \in \R^d$}.
    \item Moreover, \mbox{$D^{\top}: \R \to \R^{rd}$} denotes the \emph{upsampling operator} that acts on $d$-dimensional column vectors by subdividing each of its entries into $r$ replicas. 
\end{itemize}
\end{remark}

\subsection{Robust restoration}
\label{ssec:imagemm_robust}
\begin{figure*}[htbp]
    \centering
    \includegraphics[width=.325\textwidth]{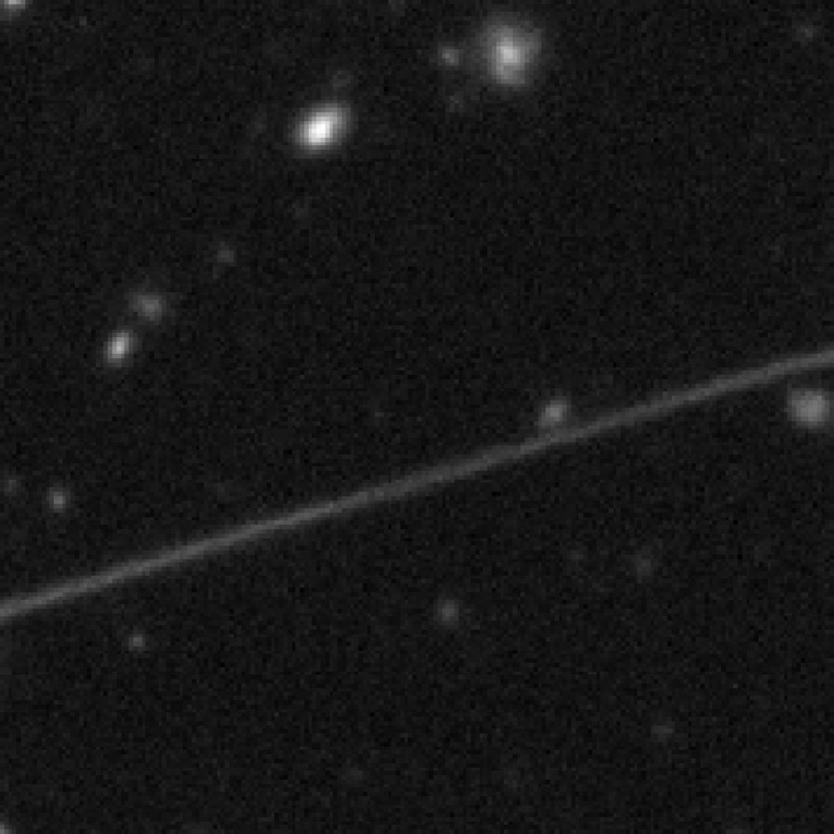}
    \includegraphics[width=.325\textwidth]{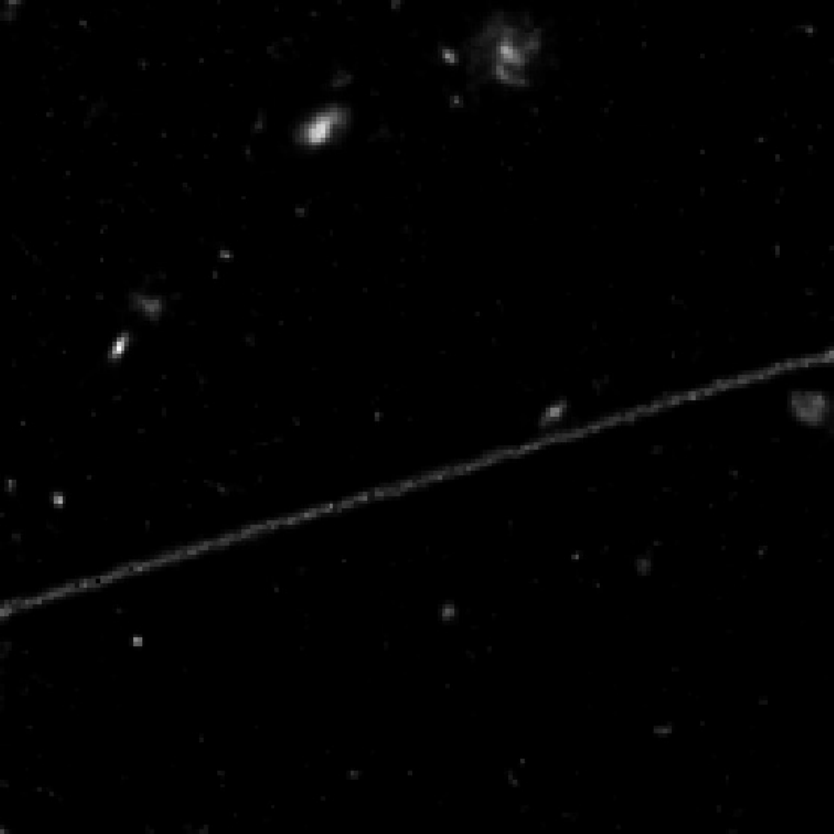}
    \includegraphics[width=.325\textwidth]{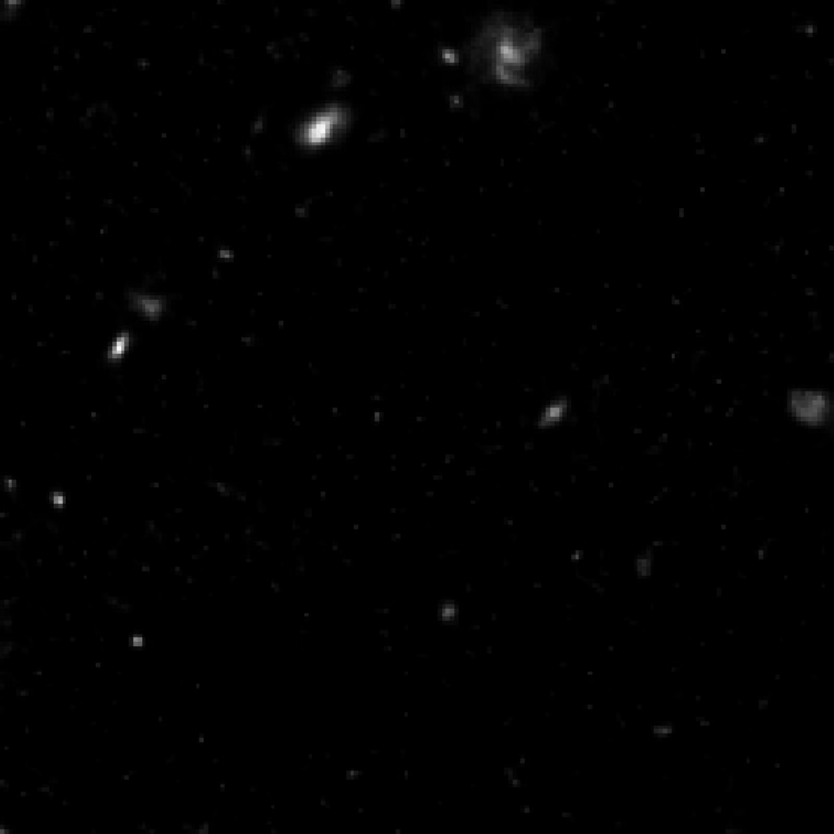}
    \caption{\textbf{Robust restoration.} \textit{Left:} Cutout from the \emph{pipeline coadd} of the HSC exposures, in which a satellite trail is present (the diagonal line through the image). \textit{Middle:} \emph{Latent image} obtained using Algorithm~\ref{alg:mm_multiframe_restoration}, in which noise levels and blur are significantly reduced in comparison to the coadd. However, Algorithm~\ref{alg:mm_multiframe_restoration} fails to entirely remove the satellite trail. 
    \textit{Right:} \emph{Robust restoration} obtained via Algorithm~\ref{alg:mm_multiframe_restoration_robust} using the Huber loss, i.e., with $\rho = H_{\delta}$, where $\delta=2$. In addition to removing noise and blur, the robust restoration procedure results in the automatic removal of the entire satellite trail.}
    \label{fig:comparison_l2vrobust}
\end{figure*}

So far, in the models of the exposures in Equations~\eqref{eq:model_exposures} and~\eqref{eq:model_exposures_super_resolution}, we have assumed that the additive noise terms follow a Gaussian distribution. However, this assumption may be violated if certain exposures contain outlier pixels. For instance, such a scenario may occur in ground-based astronomical imaging data, due to the presence of satellite trails or cosmic rays in the telescope's field of view. These extreme outliers may adversely impact the quality and interpretability of latent images of the night sky obtained from Algorithms~\ref{alg:mm_multiframe_restoration} and~\ref{alg:mm_multiframe_restoration_superresolution}, as illustrated in Figure~\ref{fig:comparison_l2vrobust}. In this section, we outline how to address this issue in order to produce latent images that are robust to outliers using the \emph{ImageMM} framework. 

We note that, for the sake of brevity, we will present our robust restoration method in the context of exposures modeled by Equation~\eqref{eq:model_exposures}, and we point out that it is straightforward to generalize the method for exposures modeled by Equation~\eqref{eq:model_exposures_super_resolution} in order to perform joint robust image restoration and super-resolution.

We start the exposition of our robust restoration method by denoting the \emph{residuals} of model~\eqref{eq:model_exposures} as
\begin{equation}
    \label{eq:residual_model_exposures}
    r_i^{(t)}(x) \doteq \frac{y_i^{(t)} - (F^{(t)}x)_i}{\sigma_i^{(t)}}
\end{equation}
for each \mbox{$i = 1,\dots, d$} and \mbox{$t=1,\dots,n$}. Intuitively, a so-called \emph{robust} restoration of the night sky $\widehat{x}$ satisfies \mbox{$r_i^{(t)}(\widehat{x}) \approx 0$} for all pixels in the exposures, \emph{except for outliers}. Our approach to produce such restorations fundamentally relies on ideas from the field of robust statistics~\citep{maronna2019robust} and is based on the method of~\citet{lee2017robust}. 

Specifically, we compute the latent image of the night sky $\widehat{x}$ as a so-called \emph{$M$-estimator}: 
\begin{equation}
    \label{eq:xhat_m_estimator}
    \widehat{x} = \underset{x \in \mathcal{X}}{\argmin} ~\frac{1}{M} \sum_{t=1}^n \sum_{i=1}^d m_i^{(t)} \, \rho \left( r_i^{(t)}(x) \right) ,
\end{equation}
where $\rho: \R \to \R$ is a \emph{robust $\rho$-function}, whose precise definition is given in Appendix~\ref{app:imagemm_mmupdaterobust_formula}. In particular, when $\rho(z) \doteq z^2 / 2$, the objective function above corresponds to the $L_2$ loss (Equation~\eqref{eq:gaussian_likelihood_square_loss}), and we thus recover the interpretation of $\widehat{x}$ as the maximum likelihood estimator for the model in Equation~\eqref{eq:model_exposures} under the Gaussian noise assumption. More generally, if a $\rho$-function satisfies \mbox{$\rho = -\log p$}, where $p$ is the probability density function for the joint distribution of pixel values of $x$ given the exposures $y$ and PSFs $f$, then the definition of $\widehat{x}$ as an $M$-estimator in Equation~\eqref{eq:xhat_m_estimator} coincides with its previous general definition as a maximum likelihood estimator in Equation~\eqref{eq:deconvolution_mle_general}. Yet, a $\rho$-function does not necessarily need to correspond to the negative log-likelihood of a given probability distribution, in which case $\widehat{x}$ is not a maximum likelihood estimator.

Formulating $\widehat{x}$ as an $M$-estimator may thus be viewed as a generalization of its previous definition as a maximum likelihood estimator. This more general formulation gives us the flexibility to choose any given $\rho$-function when computing the latent image $\widehat{x}$ via Equation~\eqref{eq:xhat_m_estimator}. In particular, we can utilize $\rho$-functions that curtail the adverse impact of outliers on $\widehat{x}$. 

An example of such a $\rho$-function is the so-called \emph{Huber loss}, denoted by $H_{\delta}: \R \to \R$, which is defined as
\begin{equation}
    \label{eq:huber_rho_function}
    H_{\delta}(z) \doteq
    \begin{cases}
        \frac{1}{2} z^2 & \text{ if } |z| \leq \delta, \\
        \delta \, \left( |z| - \frac{1}{2} \delta \right) & \text{ otherwise.}
    \end{cases}
\end{equation}

The parameter $\delta > 1$ is essentially a threshold for limiting the influence of outliers on the latent image. Indeed, when residuals are `small' (\mbox{$r_i^{(t)}(x) \leq \delta$}), the Huber loss $H_{\delta}(r_i^{(t)}(x))$ coincides with the $L_2$ loss in Equation~\eqref{eq:gaussian_likelihood_square_loss}. However, when residuals are large (\mbox{$r_i^{(t)}(x) > \delta$}), we observe that the function $H_{\delta}$ is linear. 

Notice that residuals are typically large when a pixel value $y_i^{(t)}$ is an outlier. In this scenario, the Huber loss curbs the contribution of these outliers to the overall loss by virtue of being linear, in particular when contrasted with the quadratic $L_2$ loss. By solving Equation~\eqref{eq:xhat_m_estimator} with $\rho = H_{\delta}$, with $\delta=2$ typically chosen in practice, we thus recover latent images $\widehat{x}$ that are robust to the adverse impact of such outlier pixels.

From a computational point of view, we shall once again rely on the MM algorithm to solve for robust restorations of the night sky as $M$-estimators via Equation~\eqref{eq:xhat_m_estimator}. Indeed, for any given $\rho$-function, including the Huber loss in Equation~\eqref{eq:huber_rho_function}, minimizing the objective function in Equation~\eqref{eq:xhat_m_estimator} via the MM algorithm also results in an iterative multiplicative update procedure, which possesses all of the computational benefits outlined in Section~\ref{ssec:imagemm_restoration}. Specifically, given our current guess for $\widehat{x}$ at the $k^{th}$ iteration, denoted by $x_{k-1}$, this procedure involves updating our guess as follows: 
\begin{equation*}
    x_{k} = x_{k-1} \odot u_k,
\end{equation*}
where the update image $u_k \in \R^{d+d'-1}$ is now defined as
\begin{equation}
    \label{eq:robust_update_image_def}
    u_k \doteq \frac{ \sum_{t=1}^n {F^{(t)}}{}^{\top} W_{\rho}^{(t)} y^{(t)} } { \sum_{t=1}^n {F^{(t)}}{}^{\top} W_{\rho}^{(t)} F^{(t)} x_{k-1} } ,
\end{equation}
with $W_{\rho}^{(t)}$ being the diagonal $d \times d$ matrix of \emph{robust weights} for each $t\!=\!1,\dots,n$, whose $i^{th}$ diagonal entry is
\begin{equation}
    \label{eq:robust_weights_def}
    \left( W_{\rho}^{(t)} \right)_{ii} \doteq \frac{m_i^{(t)}}{v_i^{(t)}} \ \psi \left( r_i^{(t)}(x_{k-1}) \right) ,
\end{equation}
and where $\psi: \R \to \R$ is the so-called \emph{weight function},
\begin{equation}
    \label{eq:weight_function}
    \psi(z) \doteq \frac{\rho'(z)}{z}.
\end{equation}

\begin{algorithm}[htbp]
\SetKwInOut{Input}{Input}
\SetKwInOut{Output}{Output}
\SetKwFunction{FunctionMMRestorationRobust}{ImageMMRestorationRobust}

\caption{\emph{ImageMM} algorithm for robust multi-frame astronomical image restoration.}
\label{alg:mm_multiframe_restoration_robust}

\DontPrintSemicolon

\Indm
\Input{Exposures, $y=\{y^{(1)}, \dots, y^{(n)}\}$. \\
PSFs, $f=\{f^{(1)}, \dots, f^{(n)}\}$. \\
Masks, $m=\{m^{(1)}, \dots, m^{(n)}\}$. \\
Variances $v = \{v_i^{(t)}\}$ for each pixel value $y_i^{(t)}$. \\
Robust $\rho$-function, $\rho$. \\
Initial guess for the latent image, $x_0$. \\
Maximum number of iterations, $K$. \\
Update clipping factor, $\kappa$.}
\BlankLine

\Output{Robust latent image of the night sky, $\widehat{x}$. \\
\noindent \hrulefill}

\nlnonumber
\FunctionMMRestorationRobust{$y, f, m, v, \rho, x_0, K, \kappa$}:
\BlankLine

\Indp
Initialize $\widehat{x} \gets x_0$\;
\While{$k \gets 1$ \KwTo $K$}
    {\Repeat{$u_k' \approx u_{k-1}'$}
    {\For{$t=1,\dots,n$}
    {\For{$i=1,\dots,d$}
    {$\omega_i^{(t)} \gets \frac{m_i^{(t)}}{v_i^{(t)}} \, \psi \left( r_i^{(t)}(\widehat{x}) \right)$} \;
    }
    {$W_{\rho}^{(t)} \gets \operatorname{diag}\left(\omega_1^{(t)}, \dots,  \omega_d^{(t)}\right)$}\;
    \BlankLine
    {$ u_k \gets \frac{ \sum_{t=1}^n {F^{(t)}}{}^{\top} W_{\rho}^{(t)} y^{(t)} } { \sum_{t=1}^n {F^{(t)}}{}^{\top} W_{\rho}^{(t)} F^{(t)} \widehat{x} }$\;
    \BlankLine
    $ u_k' \gets \max{\left\{ 1/ \kappa, ~\min{( \kappa, u_k)} \right\}}$\;
    \BlankLine
    $~\widehat{x} \gets \widehat{x} \odot u_k' $\;
    \BlankLine}
    }
    }
\Return{$\widehat{x}$}
\end{algorithm}

The derivations of the formulae in Equations~\eqref{eq:robust_update_image_def},~\eqref{eq:robust_weights_def} and~\eqref{eq:weight_function} are given in Appendix~\ref{app:imagemm_mmupdaterobust_formula}. In particular, when $\rho(z) = z^2/2$, we have that $\psi(z) = 1$, and we thus recover Algorithm~\ref{alg:mm_multiframe_restoration} as the definitions of the update images (Equations~\eqref{eq:update_image_def} and~\eqref{eq:robust_update_image_def}) coincide. The aforementioned robust restoration procedure may thus essentially be viewed as a generalization of Algorithm~\ref{alg:mm_multiframe_restoration}, in which the update image $u_k$ now factors in the weight function $\psi$ evaluated at the residuals $r_i^{(t)}(x_{k-1})$ at each iteration. The role of the weight function is to downweight the multiplicative update factor for those pixels in $x_{k-1}$ for which the residuals are large, i.e., for pixels that correspond to outliers in the exposures. This is precisely what limits the impact of these outliers in the resulting latent image $\widehat{x}$.

We provide a summary of our robust restoration procedure in Algorithm~\ref{alg:mm_multiframe_restoration_robust}, and illustrate results with HSC data in Figure~\ref{fig:comparison_l2vrobust}.

\section{HSC data analysis with Image-MM}
\label{sec:results}
We now provide an in-depth discussion of the implementation and computational performance of the \emph{ImageMM} framework, and present a quantitative analysis of \emph{ImageMM}'s restorations, as well as results from photometric tests performed on HSC imaging data.

\subsection{Implementation and processing times}
\label{ssec:implementation_performance}

First, we note that all algorithms in the \emph{ImageMM} framework, namely Algorithms~\ref{alg:mm_multiframe_restoration},~\ref{alg:mm_multiframe_restoration_superresolution}, and~\ref{alg:mm_multiframe_restoration_robust}, have been implemented in \texttt{TensorFlow}~\citep{abadi2016tensorflow}. Although typically utilized for the development and deployment of machine and deep learning workflows, the use of \texttt{TensorFlow} in high-performance scientific computing is now increasingly commonplace; see the computational fluid dynamics simulation framework of~\citet{wang2022tensorflow} for instance. In our case, we chose to implement \emph{ImageMM} in \texttt{TensorFlow} for the following reasons:
\begin{itemize}[leftmargin=*]
    \item \texttt{TensorFlow} is a flexible and powerful open-source library that enables rapid and modular software development, and which also contains implementations of advanced algorithmic solutions. In particular, it features a built-in implementation of the Adam optimizer, which we leverage to obtain super-resolved PSFs when solving Equation~\eqref{eq:psfs_super_resolution} as part of Algorithm~\ref{alg:mm_multiframe_restoration_superresolution}.
    \item \texttt{TensorFlow} allows users to seamlessly leverage GPU or even tensor processing unit acceleration, thus making it particularly well-suited for performing computations with high-dimensional imaging data while maintaining fast processing times. In particular, we use GPU acceleration to perform the convolutions required to compute update images (Equations~\eqref{eq:update_image_def} and~\eqref{eq:robust_update_image_def}) as part of the multiplicative update procedures in Algorithms~\ref{alg:mm_multiframe_restoration},~\ref{alg:mm_multiframe_restoration_superresolution}, and~\ref{alg:mm_multiframe_restoration_robust}. As a result, we obtain latent images of the night sky in near real time, even when processing high-resolution exposures.
\end{itemize}

\begin{table}[t]
\begin{center}
 \begin{tabular}{|| c || c | c | c ||}
    \toprule
       & $d'=25^2$ & $d'=49^2$ & $d'=99^2$  \\ [0.5ex] 
     \hline\hline
        $d=125^2$ &  $ 3.64 \times 10^{-3}$ & $8.81 \times 10^{-3}$ & $ 4.05 \times 10^{-2}$ \\
        $d=250^2$ &  $ 8.78 \times 10^{-3}$ & $ 2.53 \times 10^{-2}$ & $1.11 \times 10^{-1}$ \\ 
        $d=500^2$ &  $3.01 \times 10^{-2}$ & $ 8.89 \times 10^{-2}$ & $ 3.71 \times 10^{-1}$ \\
        $d=1000^2$ &  $1.20 \times 10^{-1}$ & $ 3.48 \times 10^{-1}$ & $1.33 \times 10^{~0}$\\
        $d=2000^2$ &  $4.75 \times 10^{-1}$ & $ 1.35 \times 10^{~0}$ & $5.05 \times 10^{~0}$ \\
     \hline
    \end{tabular}
    \caption{\textbf{Computation time table for Algorithm~\ref{alg:mm_multiframe_restoration}.} Average computation time (per iteration, in seconds) for processing cutouts of exposures with $d$ pixels, with PSFs containing $d'$ pixels, when using Algorithm~\ref{alg:mm_multiframe_restoration} to compute a latent image $\widehat{x}$ using all of the $n=33$ exposures and their corresponding PSFs from the HSC data set.}
    \label{tab:timing_alg1}
\end{center}
\end{table}

Indeed, we illustrate this latter point in Table~\ref{tab:timing_alg1}, which displays the average computation time $T$ (in seconds) required to perform one iteration of the multiplicative MM update procedure in Algorithm~\ref{alg:mm_multiframe_restoration}, which was used to obtain a latent image of the night sky $\widehat{x}$ by processing all $n=33$ images from the HSC data set. We report the computation times $T$ when processing cutouts of the exposures containing $d$ pixels, with PSFs containing $d'$ pixels, for various combinations of values of $d$ and $d'$. We note that, in Table~\ref{tab:timing_alg1}, the PSFs of size $d'=25^2$ correspond to the original HSC PSFs, in their native resolution. Meanwhile, the PSFs of size $d'=49^2$ and $d'=99^2$ were obtained with our PSF solver (Equation~\eqref{eq:psfs_super_resolution}) using super-resolution factors of $r=2$ and $r=4$ respectively, where the original HSC PSFs were used as the target PSFs. As such, we do not report computation times with Algorithm~\ref{alg:mm_multiframe_restoration_superresolution}, as it essentially involves the same computations as Algorithm~\ref{alg:mm_multiframe_restoration} but with different values of $d$ and $d'$ based on the super-resolution factor, identical to what is reported in Table~\ref{tab:timing_alg1}.

By analyzing Table~\ref{tab:timing_alg1}, we notice that, for PSFs with a fixed size of $d'$ pixels, the computation time $T$ becomes roughly $2.5-4$ times higher as the size of the exposures being processed, namely $d$, is quadrupled. Similarly, for exposures with a fixed size of $d$ pixels, the computation time $T$ becomes about $2.5-4.5$ times higher when the size of the PSFs $d'$ is quadrupled. Moreover, Table~\ref{tab:timing_alg1} gives us a sense of how rapidly one is able to obtain restored images of the night sky with \emph{ImageMM} when processing HSC data. Indeed, note that, when using cutouts containing $d=1000^2$ pixels and PSFs containing $d'=25^2$ pixels, such as the original HSC PSFs, it takes about $T=0.12s$ to perform one iteration of the multiplicative update step in Algorithm~\ref{alg:mm_multiframe_restoration}. For images and PSFs of this size, we have empirically observed that Algorithm~\ref{alg:mm_multiframe_restoration} converges in under $100$ iterations, where convergence is determined via the stopping criterion in Equation~\eqref{eq:convergence_criteria}, which is outlined in Appendix~\ref{app:stopping_criteria_implementation}. Therefore, with Algorithm~\ref{alg:mm_multiframe_restoration} of the \emph{ImageMM} framework, we obtain high-fidelity restorations $\widehat{x}$ in under $12s$ when processing exposures with $d=1000^2$ pixels and PSFs with $d'=25^2$ pixels from the HSC data set.

\begin{table}[t]
\begin{center}
 \begin{tabular}{|| c || c | c | c ||}
    \toprule
       & $d'=25^2$ & $d'=49^2$ & $d'=99^2$  \\ [0.5ex] 
     \hline\hline
        $d=125^2$ &  $ 5.73 \times 10^{-3}$ & $ 1.48 \times 10^{-2}$ & $ 7.38 \times 10^{-2}$ \\ 
        $d=250^2$ &  $ 1.55 \times 10^{-2}$ & $ 4.44 \times 10^{-2}$ & $ 1.98 \times 10^{-1}$ \\ 
        $d=500^2$ &  $ 5.50 \times 10^{-2}$ & $ 1.58 \times 10^{-1}$ & $ 6.48 \times 10^{-1}$   \\
        $d=1000^2$ &  $ 2.22 \times 10^{-1}$ & $ 6.09 \times 10^{-1}$ & $ 2.32 \times 10^{~0}$ \\
        $d=2000^2$ &  $ 8.74 \times 10^{-1}$ & $ 2.41 \times 10^{~0}$ & $ 8.82 \times 10^{~0}$  \\
     \hline
    \end{tabular}
    \caption{\textbf{Computation time table for Algorithm~\ref{alg:mm_multiframe_restoration_robust}.} Average computation time (per iteration, in seconds) for processing cutouts of exposures with $d$ pixels, with PSFs containing $d'$ pixels, when using Algorithm~\ref{alg:mm_multiframe_restoration_robust} to compute a robust latent image $\widehat{x}$ using all $n=33$ images from the HSC data set.}
    \label{tab:timing_alg3}
\end{center}
\end{table}

For the sake of completeness, we also report Table~\ref{tab:timing_alg3}, which contains the per-iteration computation times $T$ for obtaining robust restorations using the HSC data set via Algorithm~\ref{alg:mm_multiframe_restoration_robust}. We observe that, for fixed values of $d$ and $d'$, the corresponding computation times $T$ for Algorithm~\ref{alg:mm_multiframe_restoration_robust} are roughly $1.5-2$ times higher compared to those of Algorithm~\ref{alg:mm_multiframe_restoration}. This is because we recompute the matrix of robust weights (Equation~\eqref{eq:robust_weights_def}) at each iteration of Algorithm~\ref{alg:mm_multiframe_restoration_robust}, as it requires the evaluation of the weight function (Equation~\eqref{eq:weight_function}) at the residuals (Equation~\eqref{eq:residual_model_exposures}) in every iteration of the procedure. Nevertheless, the scaling in computation times as $d$ and $d'$ increase is similar to what is reported in Table~\ref{tab:timing_alg1}. In particular, Algorithm~\ref{alg:mm_multiframe_restoration_robust} takes under $22s$ to produce robust restorations using HSC exposures with $d=1000^2$ pixels and PSFs with $d'=25^2$ pixels.

\begin{remark}
We note that all experiments in this paper, including recording the computation times in Tables~\ref{tab:timing_alg1} and~\ref{tab:timing_alg3}, were performed on the \emph{SciServer} platform~\citep{taghizadeh2020sciserver} using a compute engine with an Intel Xeon Gold 6226, 12 core, 2.70 GHz CPU and a Tesla V100-SXM2 GPU.
\end{remark}

\subsection{Analysis of \emph{ImageMM}'s restoration quality}
\label{ssec:analysis_results}
We now present a quantitative assessment of the latent images produced by \emph{ImageMM}. As outlined throughout Section~\ref{sec:imagemm}, the \emph{ImageMM} framework yields high-fidelity restorations of the night sky that exhibit substantially reduced sky-background noise and enhanced source sharpness relative to the HSC pipeline coadds. This improvement in image clarity enables the recovery of fine spatial structures---such as the morphological features of galaxies---in remarkable detail.

To quantitatively evaluate the visually apparent improvements in \emph{ImageMM}'s restorations compared to the pipeline coadds, we report a set of metrics that capture the enhancement in global image sharpness and the reduction in sky-background noise. These quantitative results are summarized in Table~\ref{tab:sharpness}.

\begin{table}[htbp]
    \begin{tabular}{lcc}
        \toprule
            & \textbf{Coadd} & \textbf{\emph{ImageMM}} \\
        \hline \hline
            Sharpness, $S_F$ & 5.69 & 6.42 \\
            Noise, $\sigma_{\operatorname{sky}}$ & 8.96 $\times 10^{-2}$ & 8.50 $\times 10^{-6}$  \\
        \bottomrule
    \end{tabular}
    \caption{Quantitative comparison of sharpness and sky-background noise levels between the pipeline coadd and \emph{ImageMM}'s latent image, computed over the a field of view of size $4200 \times 4200$ pixels from the HSC survey. Higher values of $S_F$ indicate sharper images, and lower values of $\sigma_{\operatorname{sky}}$ correspond to lower background noise levels.}
\label{tab:sharpness}
\end{table}

More precisely, we quantify image sharpness by employing a Fourier-based metric, denoted $S_F$, which characterizes the high-frequency content of each image. Specifically, $S_F$ is computed by taking the logarithm of the magnitude of the two-dimensional Fourier transform of the image and averaging it over the frequency domain~\citep{krotkov1988active}. Higher values of $S_F$ correspond to a greater presence of fine-scale structures and edge detail, and are thus indicative of increased sharpness. While $S_F$ serves as an effective global sharpness measure, it may also be influenced by high-frequency noise. To account for this, we additionally report the residual sky-background noise level, $\sigma_{\operatorname{sky}}$, calculated as the standard deviation of background pixels. This computation is performed using the \texttt{sep} package, a \texttt{Python} implementation of the widely used source extraction tool \texttt{SExtractor}~\citep{bertin1996sextractor, barbary2016sep}.

As shown in Table~\ref{tab:sharpness}, the latent image produced by \emph{ImageMM} exhibits marked improvement in both sharpness and background noise suppression relative to the HSC pipeline coadd. Specifically, the sharpness metric $S_F$ increases from $5.69$ to $6.42$, corresponding to an enhancement of approximately $13\%$. Concurrently, the residual sky-background noise level, measured by $\sigma_{\operatorname{sky}}$, is reduced by over $3$ orders of magnitude---from $8.96 \times 10^{-2}$ in the coadd to $8.50 \times 10^{-6}$ in the \emph{ImageMM} restoration. These quantitative gains underscore \emph{ImageMM}'s ability to recover fine spatial structure while effectively suppressing background noise, consistent with the visual improvements observed in the figures shown in Section~\ref{sec:imagemm}. We note that all metrics reported in Table~\ref{tab:sharpness} were computed over the entire $4200 \times 4200$ pixel field of view from which the HSC exposures in Figure~\ref{fig:HSC_data} were derived, thereby ensuring that the analysis captures global image characteristics rather than localized features.

To further assess the fidelity of the restorations produced by \emph{ImageMM}, we compute two widely adopted image quality metrics: the peak signal-to-noise ratio~\citep[PSNR;][]{hore2010image} and the structural similarity index measure~\citep[SSIM;][]{wang2004image}. Specifically, we evaluate the average PSNR and SSIM values between \emph{ImageMM}'s restoration and each individual HSC exposure, as well as the PSNR and SSIM values between the \emph{ImageMM} latent image and the HSC pipeline coadd. These metrics are summarized in Table~\ref{tab:ssim_psnr}.

\begin{table}[htbp]
\small
\begin{tabular}{lcc}
\toprule
 & \multicolumn{2}{c}{\textbf{\emph{ImageMM}}} \\
\textbf{Metric} & \textbf{vs. Exposures} & \textbf{vs. Coadd} \\
\hline \hline
PSNR (dB) & 39.93 & 40.32 \\
SSIM  & 0.97 & 0.98 \\
\bottomrule
\end{tabular}
\caption{Comparison of PSNR and SSIM values between \emph{ImageMM}'s restoration versus the HSC exposures, and versus the pipeline coadd, computed across a $4200 \times 4200$ pixel field of view from the HSC survey. Higher PSNR values indicate closer numerical resemblance between the images, and higher SSIM values suggest greater perceptual fidelity between the images.}
\label{tab:ssim_psnr}
\end{table}

Table~\ref{tab:ssim_psnr} shows that \emph{ImageMM}'s restoration demonstrates high numerical and perceptual fidelity when compared to both the individual HSC exposures and the pipeline coadd. The PSNR values of $39.93$~dB (versus exposures) and $40.32$~dB (versus coadd) indicate a strong overall agreement in pixel intensities, suggesting that the latent images produced by \emph{ImageMM} are consistent with both the raw observational data and the standard coaddition product derived from these exposures. Furthermore, the SSIM values of $0.97$ and $0.98$ highlight the high perceptual similarity between the restored images and the comparison baselines.

These results, when interpreted alongside the improvements in global sharpness and the substantial reduction in background noise reported in Table~\ref{tab:sharpness}, reinforce the conclusion that \emph{ImageMM} produces high-fidelity restorations that enhance overall image quality and reveal fine-scale structure (especially in comparison to the pipeline coadd), all while remaining structurally and numerically consistent with the underlying imaging data.

\subsection{Photometric tests with HSC data using \emph{ImageMM}}
\label{ssec:photometric_results_hsc}
So far, we have seen that \emph{ImageMM} produces sharp, high-fidelity nonparametric latent images of the night sky with fast processing times, even for high-resolution data. In the resulting restorations, intricate details of bright sources are unveiled, faint sources (which are typically indistinguishable from the noisy sky background in exposures and coadds) become detectable, especially with super-resolution, and extreme outliers are effectively removed. 

Yet an important question remains: can we reliably use latent images obtained with \emph{ImageMM} in the context of scientific studies with ground-based astronomical imaging data? With a view toward answering this question, we leveraged \emph{ImageMM} to perform basic aperture photometry with HSC imaging data. 

More specifically, we compared the calibrated magnitudes of corresponding sources from the HSC pipeline coadd, and from a robust restoration of the night sky $\widehat{x}$ produced using Algorithm~\ref{alg:mm_multiframe_restoration_robust}, where we used the Huber loss in Equation~\eqref{eq:huber_rho_function} as the $\rho$-function. We note that the HSC coadd and latent image $\widehat{x}$ were derived from cutouts of HSC exposures of size \mbox{$800 \times 800$} pixels, whose center coordinates are given by R.A.~\mbox{$\alpha=150.42493^{\circ}$} and decl.~$\delta=2.06615^{\circ}$, as taken from the Sloan Digital Sky Survey (SDSS) SkyServer~\citep{szalay2002sdss}.

We carried out source detection on the coadd and our latent image $\widehat{x}$ using \texttt{sep}, the \texttt{Python} implementation of the source extraction software \texttt{SExtractor}~\citep{bertin1996sextractor, barbary2016sep}, by applying an absolute threshold of \mbox{$\tau=0.1$} on both images. Flux measurements for these extracted sources, based on aperture photometry, were performed using elliptical apertures obtained from \texttt{sep} on the un-thresholded images of the HSC coadd and $\widehat{x}$ respectively. It is important to note that a small proportion of apertures may differ between the coadd and our latent image, especially for blended resolved sources. To address this issue, we performed flux measurements by applying the same apertures (i.e., those obtained for the HSC coadd) to corresponding sources in both the coadd and the latent image. Calibrated magnitudes were then derived from these flux measurements using $i$-band magnitudes of stars from the SDSS catalog as references. 

\begin{figure*}[htbp]
    \centering
    \includegraphics[width=.935\textwidth]{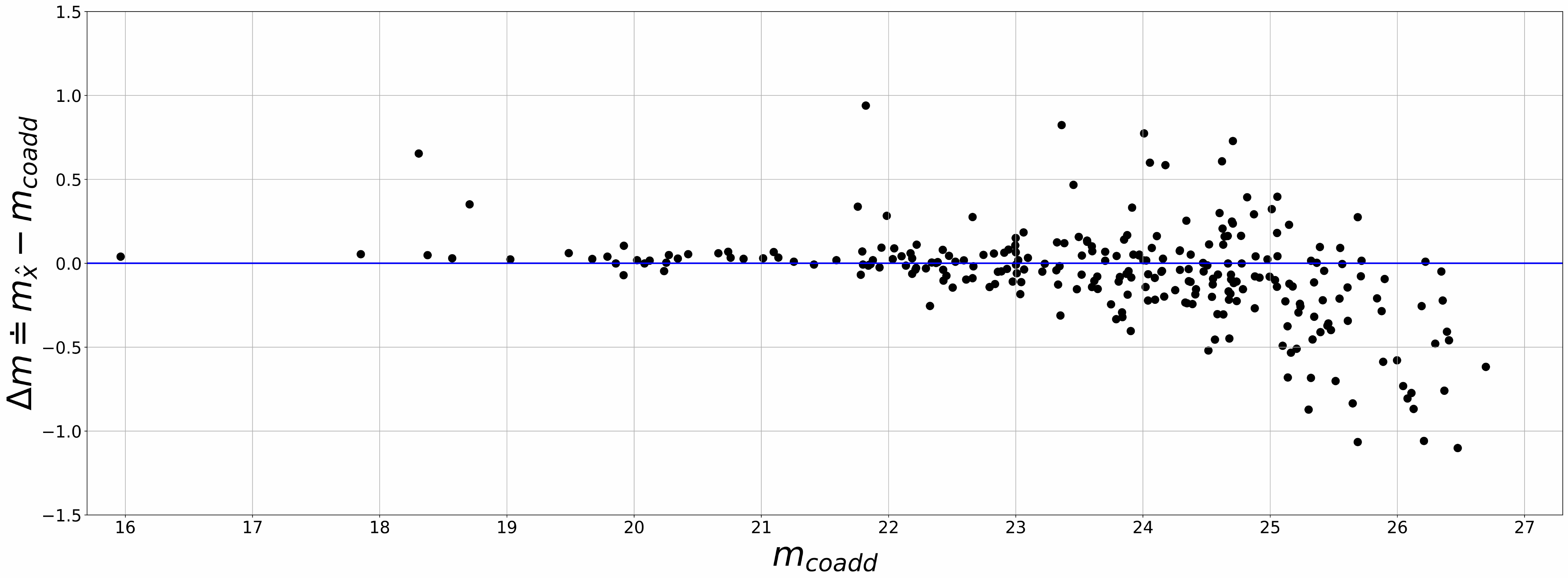}
    \caption{\textbf{Basic aperture photometry with HSC data.} The scatter plot shows the differences in magnitudes \mbox{$\Delta m \doteq m_{\widehat{x}} - m_{\operatorname{coadd}}$} as a function of \mbox{$m_{\operatorname{coadd}}$}, where $m_{\operatorname{coadd}}$ and $m_{\widehat{x}}$ are the calibrated magnitudes of corresponding sources from the HSC pipeline coadd and from \emph{ImageMM}'s latent image $\widehat{x}$, respectively. The calibrated magnitudes are consistent up to magnitudes of roughly $25$, beyond which detected sources in $\widehat{x}$ appear systematically brighter than their counterparts in the coadd.}
    \label{fig:photometric_tests_sep}
\end{figure*}

Figure~\ref{fig:photometric_tests_sep} illustrates the comparison between calibrated magnitudes of corresponding sources from the HSC pipeline coadd and from our latent image $\widehat{x}$, which we denote as $m_{\operatorname{coadd}}$ and $m_{\widehat{x}}$ respectively.

The results demonstrate that calibrated magnitudes of sources in \emph{ImageMM}'s latent image $\widehat{x}$ are consistent with those in the HSC pipeline coadd. To be more precise, the measurements are consistent up to magnitudes of roughly $25$, beyond which a small discrepancy becomes noticeable. Indeed, low-brightness sources (whose calibrated magnitudes are of roughly $25$ or more) that are detected in our latent image $\widehat{x}$ appear systematically brighter than their counterparts in the HSC coadd. This is to be expected, as \emph{ImageMM}'s restoration procedure concentrates flux from the noisy sky background into regions where sources are located in the resulting latent image. This effect is more pronounced for faint objects, as depicted in Figure~\ref{fig:lbo_restoration}, thereby explaining the deviation in calibrated magnitudes for low-brightness objects observed in Figure~\ref{fig:photometric_tests_sep}. 

These results show that, relative to pipeline coadds, \emph{ImageMM} yields latent images in which the fluxes of bright sources are preserved, and in which faint objects appear systematically brighter. In particular, the findings demonstrate that \emph{ImageMM} is a viable framework for obtaining physically meaningful restorations of the night sky behind the atmosphere, which are comparable to the pipeline coadds that are typically used in the context of scientific studies with cutting-edge astronomical imaging data. Additional evidence to corroborate these observations is provided in Appendix~\ref{app:imagemm_simulated_data}, where \emph{ImageMM} was evaluated on simulated exposures.

\section{Conclusion and outlook}
\label{sec:conclusion_astroimage}
To conclude, we have introduced \emph{ImageMM}, a new framework based on the MM algorithm for multi-frame astronomical image restoration and super-resolution. Our method overcomes varying levels of atmospheric blur in astronomical exposures to produce clear, high-fidelity, nonparametric latent images of the night sky.

A defining feature of \emph{ImageMM} is the iterative multiplicative MM update procedure used to produce restorations, in which all exposures are processed simultaneously at each step (see Section~\ref{ssec:imagemm_restoration}). In sharp contrast to existing streaming methods, this results in latent images that do not depend on the order in which exposures are processed. 

Moreover, our framework features a novel variational approach for computing refined PSF estimates of arbitrary resolution, which are required for super-resolution, and which are particularly well-adapted for astronomical image processing (see Section~\ref{ssec:super_resolution}).

We tested \emph{ImageMM} on imaging data from the HSC survey. In particular, our method is able to effectively deblur a wide range of objects in the HSC exposures, including extended sources such as galaxies, as well as point sources such as stars. Indeed, in latent images obtained using the \emph{ImageMM} framework, we successfully recover sharp details such as the intricate shapes of spiral galaxy arms, especially when super-resolution is applied (see Figure~\ref{fig:galaxy_restoration}). Moreover, small, faint objects also become detectable in the super-resolved latent images, even at subpixel scales (see Figure~\ref{fig:lbo_restoration}), due to the use of the refined PSF estimates in the restoration procedure (see Figure~\ref{fig:psf_solver}). In particular, the results suggest that super-resolved restorations produced with \emph{ImageMM} have the potential to expand our source detection limits. 

The flexibility of the \emph{ImageMM} framework also allows us to incorporate robust statistical estimation into the restoration process, which allows us to automatically remove extreme outliers, such as satellite trails, in the latent images (see Figure~\ref{fig:comparison_l2vrobust}).

Furthermore, from a computational point of view, the \texttt{TensorFlow} implementation of \emph{ImageMM} enables us to leverage GPU acceleration when performing the iterative multiplicative MM updates in Algorithms~\ref{alg:mm_multiframe_restoration},~\ref{alg:mm_multiframe_restoration_superresolution}, and~\ref{alg:mm_multiframe_restoration_robust}, thereby yielding latent images of the night sky in near real time, even when processing high-resolution exposures (see Tables~\ref{tab:timing_alg1} and~\ref{tab:timing_alg3}).

In addition, a quantitative analysis of \emph{ImageMM}'s latent images shows that it yields restorations with enhanced sharpness and suppressed background noise, while maintaining structural and numerical consistency with the underlying data (see Section~\ref{ssec:analysis_results}). Furthermore, preliminary photometric tests demonstrate consistent source detection and calibrated magnitude measurements in the pipeline coadd of the HSC exposures and in \emph{ImageMM} restorations (see Figure~\ref{fig:photometric_tests_sep}). This indicates that our framework produces high-fidelity latent images of the night sky that are comparable to the coadds that are typically used in practice for scientific studies with ground-based astronomical data. 

These highly encouraging results demonstrate that \emph{ImageMM} is a viable framework for use in the data processing and analysis pipelines of modern ground-based astronomical surveys, such as the Hyper Suprime-Cam survey and the upcoming LSST from the Rubin Observatory. 

Building on this work, one can envisage a plethora of applications using the latent images obtained with \emph{ImageMM}. These include a number of novel photometric studies, which are currently underway. Furthermore, \emph{ImageMM} could be integrated with and benchmarked against state-of-the-art deep learning models for image restoration---such as \emph{Pix2PixHD}~\citep{wang2018high}, \emph{DDPM}~\citep{ho2020denoising}, and \emph{AstroClearNet}~\citep{sukurdeep2025astroclearnet}---to further enhance its capabilities and enable advanced analyses of cutting-edge astronomical imaging data. Moreover, given the virtual absence of sky-background noise in \emph{ImageMM} restorations, the extraction of high-quality geometric data via appropriate segmentation techniques and photometric tools becomes feasible. This paves the way for shape analysis applications, such as supernova classification or automated morphology-based galaxy classification, using frameworks for the shape analysis of curves~\citep{srivastava2010shape, bauer2017numerical, bauer2024elastic} and shape graphs~\citep{sukurdeep2022new, bal2024statistical}.

\section*{Acknowledgements}
The authors thank Yusra AlSayyad for providing access to the Hyper Suprime-Cam imaging data, and for extending valuable support with regards to the use of the LSST Science Pipelines for processing the data. 

Y.S. gratefully acknowledges support from the NVIDIA Academic Hardware Grant Program. 

T.B. gratefully acknowledges support from the National Science Foundation (Award 1909709 and Award 2206341). 

A.J.C. gratefully acknowledges support from the U.S. Department of Energy, Office of Science (Award DE-SC0011665).

\appendix

\section{MM update formula for $L_2$ loss}
\label{app:imagemm_mmupdatel2_formula}
We derive the MM update formulae given in Equations~\eqref{eq:mm_update_l2_loss} and~\eqref{eq:update_image_def}. Recall the $L_2$ loss formula (Equation~\eqref{eq:gaussian_likelihood_square_loss}), which can be expressed as follows:
\begin{equation}
    \label{eq:l2_loss_expandedform}
    \mathcal{L}(x\!\mid y, f ) = \frac{1}{2M} \sum_{t=1}^n \left\{ y^{(t)}{}^{\top} W^{(t)}y^{(t)} - 2y^{(t)}{}^{\top} W^{(t)} F^{(t)}x + x{}^{\top}F^{(t)}{}^{\top} W^{(t)} F^{(t)}x \right\},
\end{equation}
where $W^{(t)}$ is a diagonal $d \times d$ matrix with diagonal entries $W^{(t)}_{ii} \doteq m_i^{(t)} / ( \sigma_i^{(t)} )^2$ for all $i=1,\dots,d$ and $t\!=\!1,\dots,n$. This $L_2$ loss is majorized by an auxiliary function (Equation~\eqref{eq:auxiliary_l2_loss}), which we rewrite below:
\begin{equation*}
    \ell(x \mid \Tilde{x}) = \frac{1}{2M} \sum_{t=1}^n \left\{ y^{(t)}{}^{\top} W^{(t)}y^{(t)} - 2y^{(t)}{}^{\top} W^{(t)} F^{(t)}x + \Tilde{x}{}^{\top}F^{(t)}{}^{\top} W^{(t)} F^{(t)} \left( \frac{x \odot x}{\Tilde{x}} \right) \right\}.
\end{equation*}
The gradient of this auxiliary function (with respect to $x$) is given by
\begin{equation}
    \label{eq:gradient_l2_auxiliaryfunction}
    \nabla_x~\!\ell(x \mid \Tilde{x}) = \frac{1}{M} \sum_{t=1}^n \left\{ -F^{(t)}{}^{\top} W^{(t)} y^{(t)} + \left( F^{(t)}{}^{\top} W^{(t)} F^{(t)} \Tilde{x} \right) \odot \left( \frac{x}{\Tilde{x}} \right) \right\},
\end{equation}
where we have used the fact that $W^{(t)} = W^{(t)}{}^{\top}$, as it is a diagonal matrix, and $\nabla_x \left( c^{\top} \left(x \odot x \right) \right) = 2 ~c \odot x$ for all \mbox{$c, x \in \R^d$} (for any $d$) to write the second term in the expression above. 

Next, we view $\tilde x = x_{k-1}$ as the current guess for the latent image $\widehat{x}$ (at the $k^{th}$ iteration). Per the update rule (Equation~\eqref{eq:mm_step}), the updated guess $x_k$ is required to be a minimizer, and hence a stationary point, of the auxiliary function $\ell(x \mid x_{k-1})$. Therefore, the updated guess $x_k$ satisfies the first-order optimality condition \mbox{$\nabla_x~\!\ell(x \mid x_{k-1}) \big|_{x=x_k}\! = 0$}. Using the expression for the gradient of $\ell$ from Equation~\eqref{eq:gradient_l2_auxiliaryfunction}, this optimality condition simplifies to the following equation:
\begin{equation}
    \label{eq:foc_l2_auxiliaryfunction}
     \left( \frac{x_k}{x_{k-1}} \right) \odot \left( \sum_{t=1}^n  F^{(t)}{}^{\top} W^{(t)} F^{(t)} x_{k-1} \right)  = \sum_{t=1}^n F^{(t)}{}^{\top} W^{(t)} y^{(t)} .
\end{equation}
By solving Equation~\eqref{eq:foc_l2_auxiliaryfunction} for $x_k$, we obtain the closed-form expression of the MM update formulae in Equations~\eqref{eq:mm_update_l2_loss} and~\eqref{eq:update_image_def}.

\section{MM update formula for robust $\rho$-function}
\label{app:imagemm_mmupdaterobust_formula}
We now derive MM update formulae (Equations~\eqref{eq:robust_update_image_def},~\eqref{eq:robust_weights_def}, and~\eqref{eq:weight_function}) for the robust reconstruction procedure presented in Section~\ref{ssec:imagemm_robust}. First, let us recall the residuals (Equation~\eqref{eq:residual_model_exposures}) of the model for the exposures in Equation~\eqref{eq:model_exposures}, which we rewrite as follows:
\begin{equation}
    \label{eq:residuals_model_exposures_expanded}
    r_i^{(t)}(x) \doteq \frac{y_i^{(t)} - \sum_j F^{(t)}_{ij} x_j}{\sigma_i^{(t)}} \quad \text{for all \mbox{$i = 1,\dots, d$} and \mbox{$t=1,\dots,n$}.}
\end{equation}
Let $x_{\ell}$ denote the $\ell^{th}$ coordinate of the latent image $x$. For all $\ell$, the derivative of the residual \mbox{$r_i^{(t)}(x)$} with respect to $x_{\ell}$ is given by
\begin{equation}
    \label{eq:residuals_derivative_wrt_x}
    \frac{\partial}{\partial x_{\ell}} \left(r_i^{(t)}(x)\right) = -\frac{1}{\sigma_i^{(t)}} \sum_j F_{ij}^{(t)} \delta_{j \ell} = - \frac{F_{i \ell}^{(t)}}{\sigma_i^{(t)}}  ,
\end{equation}
where $\delta_{j \ell}$ is the Kronecker delta function. This derivative will be useful for the ensuing derivations.

Next, let us recall the definition of a so-called $\rho$-function, which refers to a function $\rho: \R \to \R$ satisfying the following properties~\citep[Chapter 2.3]{maronna2019robust}:
\begin{itemize}
    \item[1.] $\rho(z)$ is a nondecreasing function $|z|$.
    \item[2.] $\rho(0) = 0$.
    \item[3.] $\rho(z)$ is increasing for $z > 0$ such that $\rho(z) < \rho(+\infty)$, with $\rho(+\infty)=1$ if $\rho$ is bounded. 
\end{itemize}

These $\rho$-functions are central to our robust reconstruction approach, which involves computing the latent image of the night sky $\widehat{x}$ as a so-called \emph{$M$-estimator} for the model from Equation~\eqref{eq:model_exposures}, namely 
\begin{equation}
    \label{eq:xhat_m_estimator_appendix}
    \widehat{x} = \underset{x \in \mathcal{X}}{\argmin} ~\frac{1}{M} \sum_{t=1}^n \sum_{i=1}^d ~\! m_i^{(t)} \, \rho \left( r_i^{(t)}(x) \right) ,
\end{equation}
where $\rho: \R \to \R$ is a given $\rho$-function, such as the Huber loss (Equation~\eqref{eq:huber_rho_function}). 

We seek a way to solve the problem above via the MM algorithm. To do so, we start by noting that, since $\widehat{x}$ minimizes the objective function above, we know that $\widehat{x}_{\ell}$ satisfies the following first-order optimality condition:

\begin{equation}
    \label{eq:foc_rho_function}
    \frac{\partial}{\partial x_{\ell}} \left( \sum_{t=1}^n \sum_{i=1}^d ~\! m_i^{(t)} \rho \, \left( r_i^{(t)}(\widehat x) \right) \right) = 0 \quad \text{for all $\ell$.}
\end{equation}

By using the chain rule, together with the definition of the residuals in Equation~\eqref{eq:residuals_model_exposures_expanded} and the expression for their derivatives given in Equation~\eqref{eq:residuals_derivative_wrt_x}, one can show that the first-order condition above can be expressed as follows:

\begin{equation}
    \label{eq:foc_rho_function_with_weights}
    \sum_{t=1}^n \sum_{i=1}^d F_{\ell i}^{(t)^{\top}} \cdot w_i^{(t)} \cdot \Big( y_i^{(t)} - \sum_j F^{(t)}_{ij} \widehat{x}_j \Big) = 0 \quad \text{for all $\ell$,}
\end{equation}
where $w_i^{(t)}$ are \emph{weights} that are defined as
\begin{equation}
    \label{eq:robust_weights_for_x_update}
    w_i^{(t)} \doteq \frac{m_i^{(t)}}{( \sigma_i^{(t)} )^2} \cdot \psi \left(r_i^{(t)}(\widehat x) \right) \quad \text{for all $i=1,\dots,d$ and $t=1,\dots,n$,}
\end{equation}
and where $\psi: \R \to \R$ is the so-called \emph{weight function} previously defined in Equation~\eqref{eq:weight_function}, namely
\begin{equation}
    \label{eq:weight_function_appendix}
    \psi(z) \doteq \frac{\rho'(z)}{z} .
\end{equation}
Using the weights defined in Equation~\eqref{eq:robust_weights_for_x_update}, let us define the diagonal $d \times d$ \emph{matrix of robust weights}:
\begin{equation}
    \label{eq:matrix_robust_weights}
    W_{\rho}^{(t)} = \operatorname{diag}\Big(w_{1}^{(t)}, \dots, w_{d}^{(t)}\Big) \quad \text{for each $t=1,\dots,n$.}
\end{equation}
Using these matrices, we can rewrite the first-order condition from Equation~\eqref{eq:foc_rho_function_with_weights} in matrix form as follows:
\begin{equation}
    \label{eq:foc_rho_function_with_weights_matrixform}
\sum_{t=1}^n \sum_{i=1}^d F_{\ell i}^{(t)^{\top}} \cdot w_{i}^{(t)}\cdot \Big( y_i^{(t)} - \sum_j F^{(t)}_{ij} \widehat{x}_j \Big) = 0 \text{ 
  for all $\ell$} \quad \iff  \quad \sum_{t=1}^n F^{(t)^\top} W_{\rho}^{(t)} y^{(t)} = \sum_{t=1}^n F^{(t)^\top} W_{\rho}^{(t)} F^{(t)} \widehat{x}
\end{equation}
Notice that, if we fix the matrices of robust weights $W_{\rho}^{(t)}$, then (the matrix form of) Equation~\eqref{eq:foc_rho_function_with_weights_matrixform} corresponds to the first-order optimality condition for the following \emph{weighted $L_2$ loss}:
\begin{equation}
    \label{eq:weighted_l2loss_for_rholoss}
    \mathcal{L_{\rho}}(x\!\mid y, f ) = \frac{1}{2M} \sum_{t=1}^n \left\{ y^{(t)}{}^{\top} W_{\rho}^{(t)}y^{(t)} - 2y^{(t)}{}^{\top} W_{\rho}^{(t)} F^{(t)}x + x{}^{\top}F^{(t)}{}^{\top} W_{\rho}^{(t)} F^{(t)}x \right\}.
\end{equation}
In other words, we have shown that the robust reconstruction of the night sky that we are after, namely the $M$-estimator $\widehat{x}$ that minimizes the $\rho$-function loss in Equation~\eqref{eq:xhat_m_estimator_appendix}, turns out to also be a minimizer of the weighted $L_2$ loss in Equation~\eqref{eq:weighted_l2loss_for_rholoss}. Therefore, we can solve Equation~\eqref{eq:xhat_m_estimator_appendix} by instead minimizing the weighted $L_2$ loss in Equation~\eqref{eq:weighted_l2loss_for_rholoss}. 

In particular, we can utilize the MM algorithm to do so. By performing the same calculations as in Appendix~\ref{app:imagemm_mmupdatel2_formula}, we can show that minimizing Equation~\eqref{eq:weighted_l2loss_for_rholoss} via the MM algorithm eventually boils down to solving the following equation:
\begin{equation}
    \label{eq:foc_weightedl2_auxiliaryfunction}
     \left( \frac{x_k}{x_{k-1}} \right) \odot \left( \sum_{t=1}^n  F^{(t)}{}^{\top} W_{\rho}^{(t)} F^{(t)} x_{k-1} \right)  = \sum_{t=1}^n F^{(t)}{}^{\top} W_{\rho}^{(t)} y^{(t)} ,
\end{equation}
where $W_{\rho}^{(t)}$ are the matrices of robust weights  defined in Equation~\eqref{eq:matrix_robust_weights}, whose diagonal entries are given by the weights in Equation~\eqref{eq:robust_weights_for_x_update}, in which the weight function $\psi$ is evaluated at the residuals at the current iterate $x_{k-1}$, namely
\begin{equation*}
    w_i^{(t)} \doteq \frac{m_i^{(t)}}{( \sigma_i^{(t)} )^2} \  \psi \left(r_i^{(t)}( x_{k-1}) \right) \quad \text{for all $i=1,\dots,d$ and $t=1,\dots,n$.}
\end{equation*}
Solving Equation~\eqref{eq:foc_weightedl2_auxiliaryfunction} explicitly for $x_k$ gives rise to the closed-form expressions for the MM update formulae (Equations~\eqref{eq:robust_update_image_def},~\eqref{eq:robust_weights_def}, and~\eqref{eq:weight_function}) from Section~\ref{ssec:imagemm_robust}.

\section{Implementation of stopping criterion}
\label{app:stopping_criteria_implementation}

The convergence criterion for Algorithms~\ref{alg:mm_multiframe_restoration},~\ref{alg:mm_multiframe_restoration_superresolution}, and~\ref{alg:mm_multiframe_restoration_robust} involves determining when the clipped update images stop fluctuating. More precisely, we terminate the iterative multiplicative MM update procedure at the $k^{th}$ iteration if $u'_k \approx u'_{k-1}$, as outlined in Section~\ref{ssec:imagemm_restoration}.

A computationally tractable necessary condition for determining if $u_k' \approx u_{k-1}'$ is to check that
\begin{equation}
        \label{eq:convergence_criteria}
        \left| \frac{1}{\widetilde{M}} \sum_i \left( \Tilde{m}~\odot~\xi_k \right)_i  - 1 \right| < \varepsilon,
    \end{equation}
where the terms are defined as follows:
\begin{itemize}
    \item $\xi_k \doteq u'_k / u'_{k-1}$ is the ratio of the successive clipped update images at the $k^{th}$ iteration.
    \item $\Tilde{m}$ is the \emph{effective mask}, which is a binary-valued array of the same resolution as the reconstruction $\widehat{x}$, whose entries---for a given value of $\mu \in (0,1)$---are defined as follows:
    \begin{equation*}
    \Tilde{m}_i \doteq 
    \begin{cases} 
    1, \quad \text{if ${\left(\frac{1}{n}\sum_{t=1}^n F^{(t)}{}^{\top} m^{(t)} \right)}_i > \mu$,} \\
    0, \quad \text{otherwise.}
    \end{cases}
    \end{equation*}
    To understand the definition above, consider a given pixel $i$. If the proportion of corresponding mask entries $\{ m_i^{(t)} \}_{t=1}^n$ satisfying $m_i^{(t)} = 1$ is greater than $\mu$, then we consider pixel $i$ to be `effective' for use in the reconstruction, and encode this information in the effective mask by setting $\Tilde{m}_i = 1$. 
    \item $\widetilde{M} \doteq \sum_i \tilde{m}_i$ is the total number of effective pixels for the reconstruction.
    \item $\epsilon > 0$ is a small tolerance parameter.
\end{itemize}
In other words, enforcing the stopping condition $u'_k \approx u'_{k-1}$ via the criterion in Equation~\eqref{eq:convergence_criteria} involves checking whether, on average, the ``effective" entries of the ratio of successive clipped update images at the $k^{th}$ iteration are roughly equal to $1$, up to some tolerance parameter $\varepsilon$.
    
In the implementations of Algorithms~\ref{alg:mm_multiframe_restoration},~\ref{alg:mm_multiframe_restoration_superresolution}, and~\ref{alg:mm_multiframe_restoration_robust} that were used to produce the results in this paper, we enforced the stopping criterion via Equation~\eqref{eq:convergence_criteria}, and we typically used values of $\mu = 0.1$ to construct the effective mask, with a stopping tolerance of $\varepsilon \in (10^{-4}, 10^{-6})$.

\section{Application of Image-MM to simulated data}
\label{app:imagemm_simulated_data}
Throughout this paper, we have demonstrated results that highlight the performance and capabilities of \emph{ImageMM} on real ground-based astronomical imaging data from the Hyper Suprime-Cam survey (see Figure~\ref{fig:HSC_data}). To complement these findings and provide further validation for \emph{ImageMM}'s capabilities, we now present results obtained by deploying \emph{ImageMM} on simulated data.

To produce simulated observations of the night sky, we employed the \texttt{GalSim} simulation toolkit~\citep{rowe2015galsim} to generate high-fidelity synthetic exposures that emulate the imaging characteristics of ground-based surveys, such as the Subaru Telescope’s $i$-band observations~\citep{aihara2018hyper}. Each simulated exposure was constructed by combining both stellar and galactic sources. Galaxies were modeled using a combination of S\'{e}rsic profiles, representing elliptical galaxies, and a composite bulge-plus-disk structure for spirals. The positions and intrinsic properties of the galaxies (i.e., flux, size, and morphology) were drawn randomly from uniform distributions to mimic the diversity of real extragalactic fields. Stellar sources were simulated as point-like objects with fluxes derived from a power-law magnitude distribution (skewed toward fainter sources), convolved with synthetic PSFs. A different PSF was generated for each exposure, with full width at half maximum values drawn from a uniform range of $0.5$ to $1.0$ arcseconds and randomly selected between Kolmogorov and Gaussian profiles to capture variability in observational conditions. Sky background levels were determined based on typical $i$-band surface brightnesses for dark-sky sites (i.e., magnitudes of $20-22$) and incorporated alongside both Poisson noise and Gaussian read noise to replicate photon shot noise and detector electronics. Bad pixels were artificially introduced at random locations across the detector grid to emulate instrumental defects.

\begin{figure}[htbp]
    \centering
    \includegraphics[width=.3\textwidth]{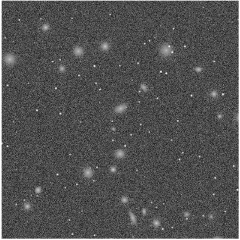}
    \includegraphics[width=.3\textwidth]{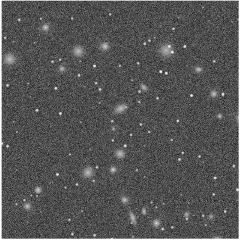}
    \includegraphics[width=.3\textwidth]{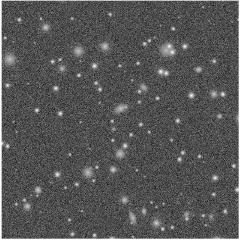}
    \includegraphics[width=.3\textwidth]{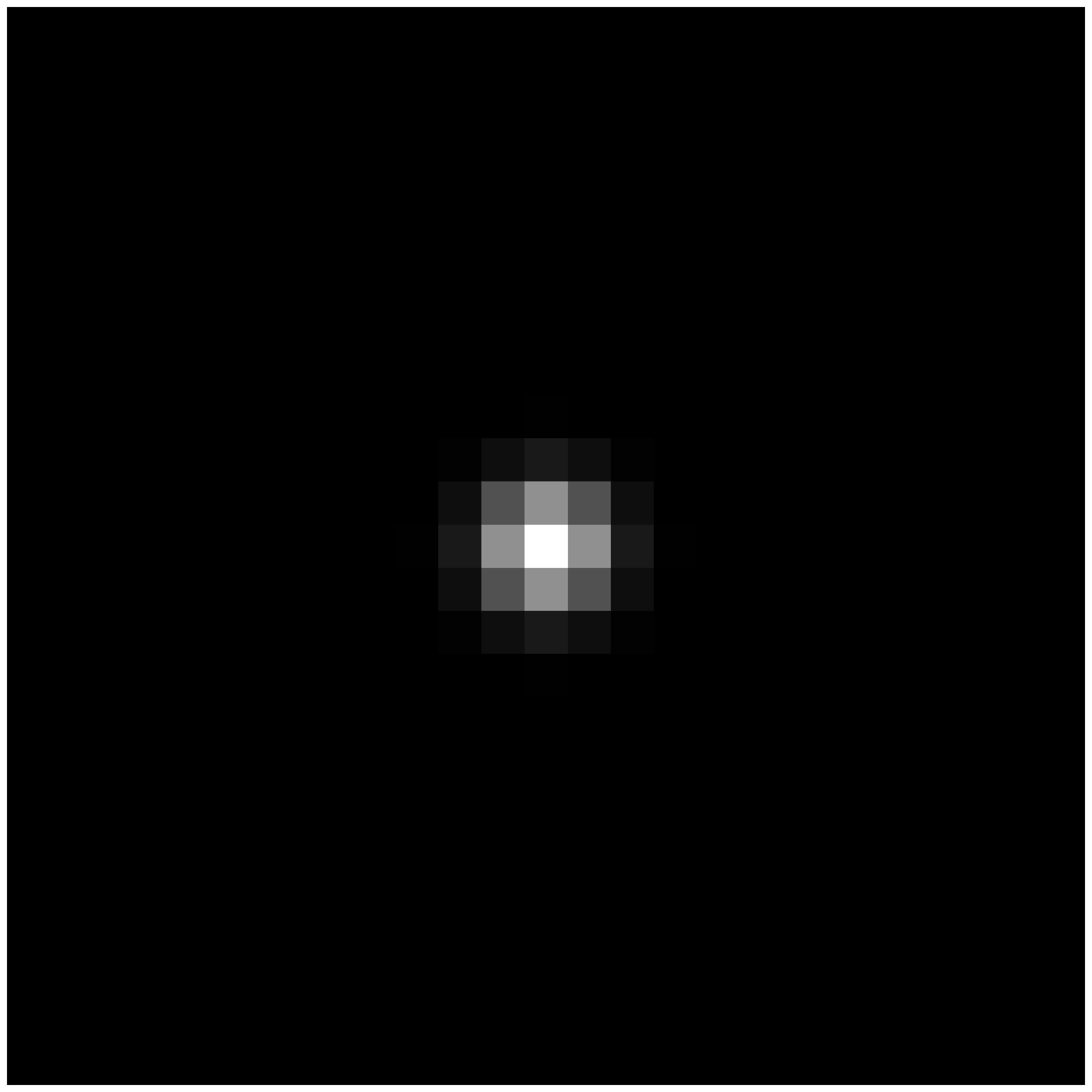}
    \includegraphics[width=.3\textwidth]{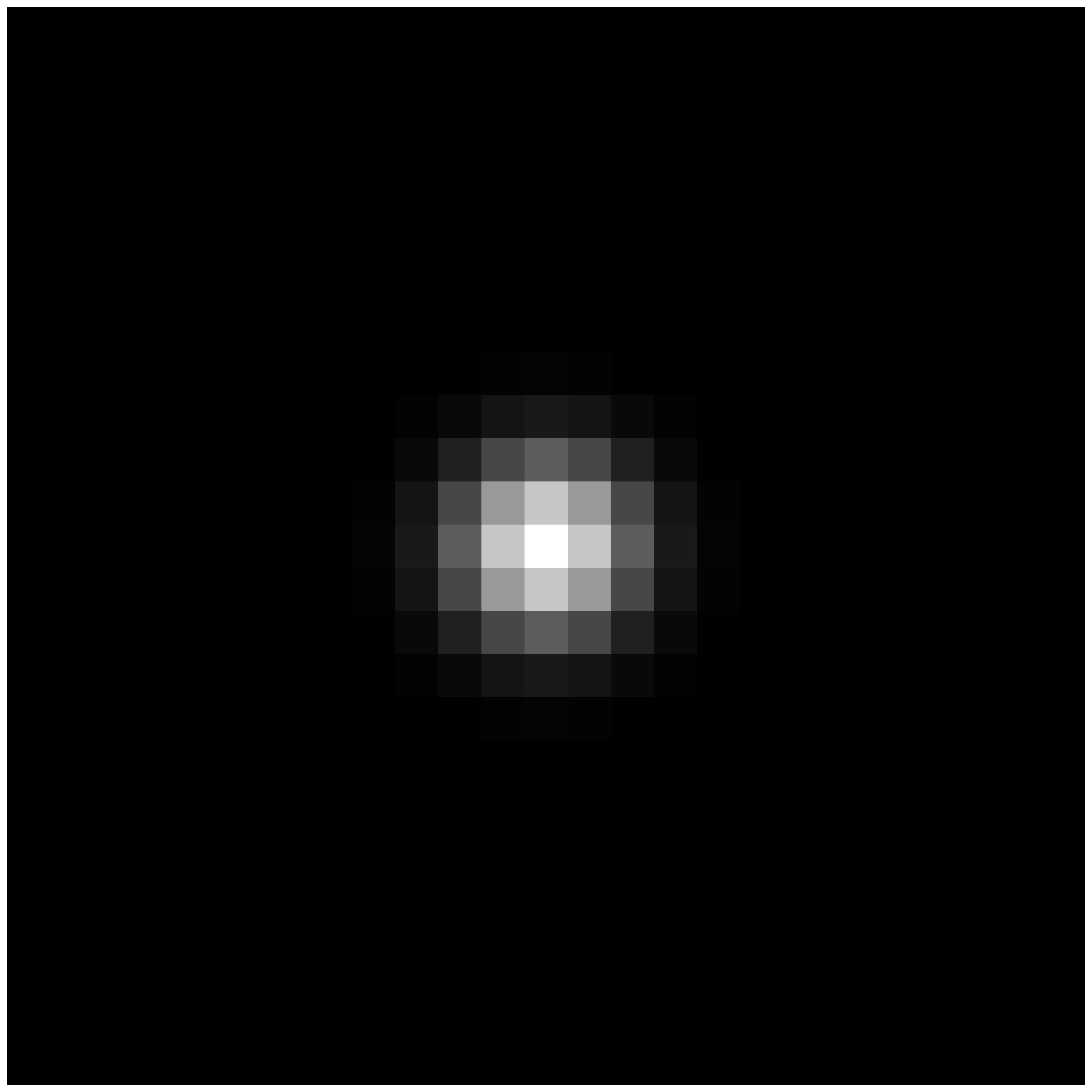}
    \includegraphics[width=.3\textwidth]{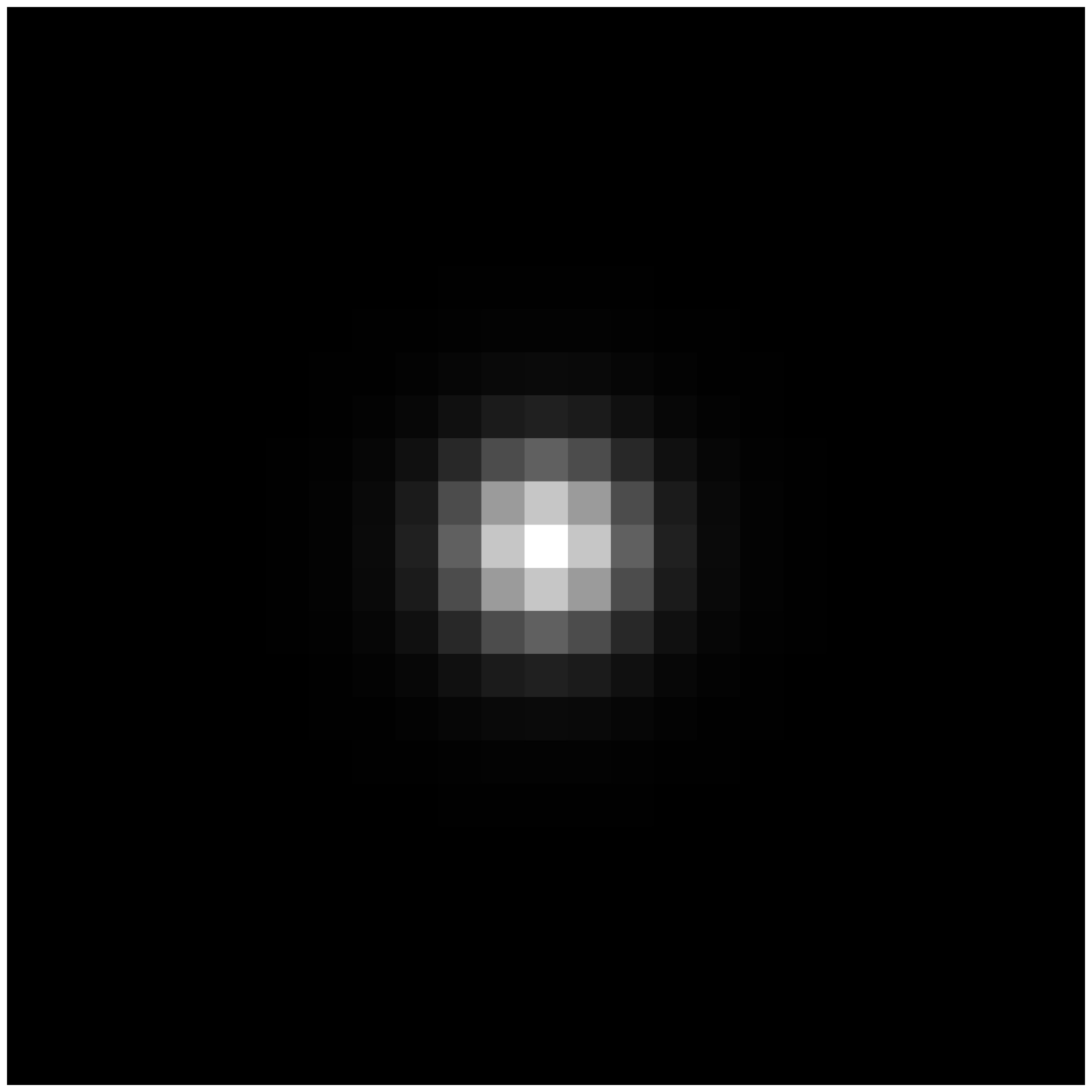}
    \caption{\textbf{Simulated imaging data.} \textit{Top row:} Selected cutouts of size $800 \times 800$ pixels from simulated exposures of the night sky, which have had the sky background subtracted. These exposures all depict the same field of view, but have different levels of noise and atmospheric blur. \textit{Bottom row:} The corresponding PSFs of size $25 \times 25$ pixels used to model atmospheric blur for each of the simulated exposures in the top row (not to scale).}
    \label{fig:galsim_data}
\end{figure}

Our simulated data set is shown in Figure~\ref{fig:galsim_data}. While it is admittedly less realistic and less complex than real observations, such as those from the HSC survey, it nevertheless serves as a useful and controlled testbed for the evaluation and preliminary validation of image processing pipelines.

\begin{figure*}[htbp]
    \centering
    \includegraphics[width=0.3\textwidth]{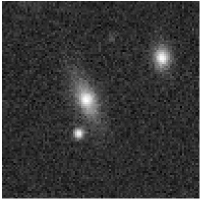}
    \includegraphics[width=0.3\textwidth]{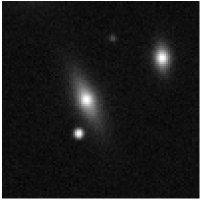}
    \includegraphics[width=0.3\textwidth]{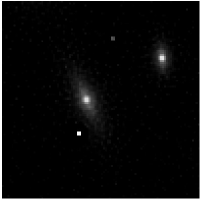}
        \caption{\textbf{Restoration of simulated observations.} \textit{Left:} Cutout of a \emph{simulated exposure} from a faint region of the sky containing multiple sources, in which the sky background is noisy. \textit{Middle:} \emph{Coadd} of the simulated exposures, in which sky-background noise is reduced, but where the sources are still blurry. \textit{Right:} Our \emph{latent image} $\widehat{x}$ computed with Algorithm~\ref{alg:mm_multiframe_restoration}. There is significantly less background noise in the restoration $\widehat{x}$, and the sources appear significantly sharper than in the coadd.} 
        \label{fig:galsim_results}
\end{figure*}

With this simulated data in hand, we applied the \emph{ImageMM} framework to perform multi-frame image restoration. Our results are displayed in Figure~\ref{fig:galsim_results}, where we focus on a faint region of the sky containing a blend of point sources and extended sources. We compare a simulated exposure, a traditional coadd produced from the simulated exposures, and the latent image obtained using Algorithm~\ref{alg:mm_multiframe_restoration}. These results once again show that \emph{ImageMM} produces a restoration in which almost all background noise from the exposures is removed, and where atmospheric blur is significantly reduced (thus increasing overall levels of sharpness in the image), especially when compared to the traditional coadd. In particular, these results are consistent with those observed on real data reported in Figures~\ref{fig:x_hat_iterates},~\ref{fig:galaxy_restoration}, and~\ref{fig:lbo_restoration}, and with our quantitative analysis from Section~\ref{ssec:analysis_results}.

We note that, while \emph{ImageMM} is very effective at removing sky-background noise, it does not result in restorations that are entirely \emph{noise-free}. This can be seen in Figure~\ref{fig:galsim_results}, where sources in the latent image are very easily discernible, but where a few pixels in the sky background have a small nonzero intensity representing noise, especially around bright extended sources.

\begin{figure*}[htbp]
    \centering
    \includegraphics[width=.85\textwidth]{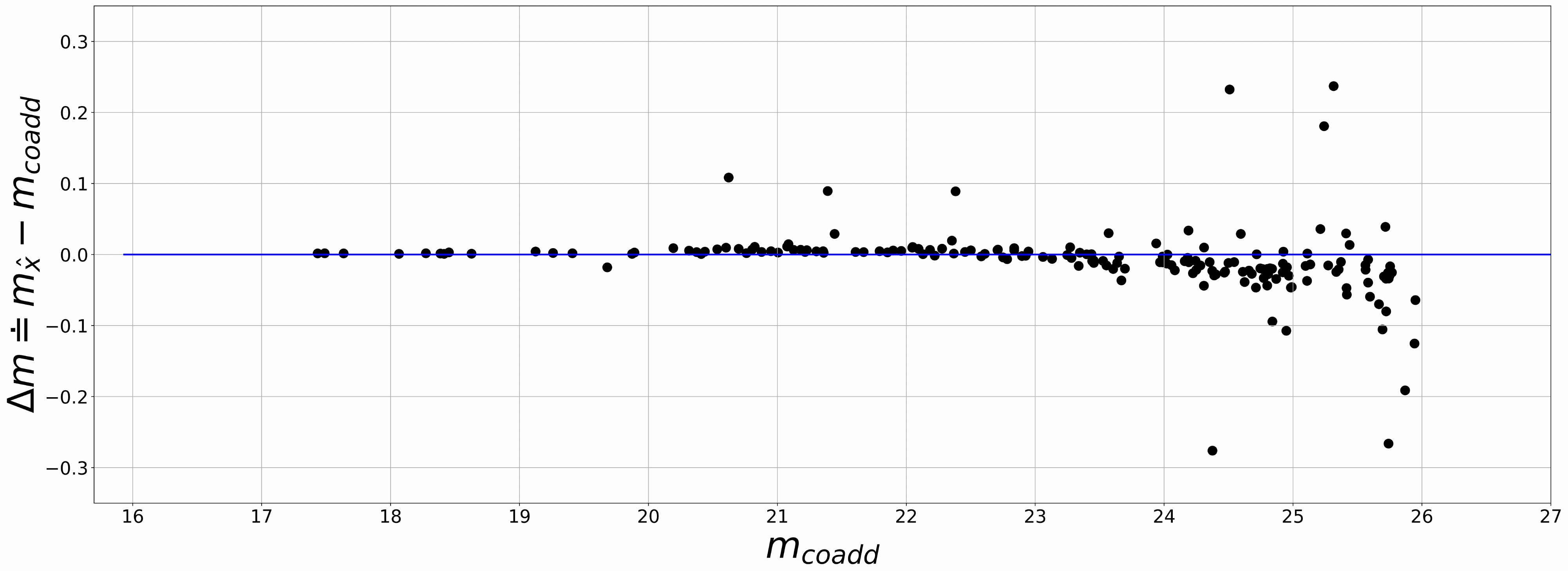}
    \caption{\textbf{Basic aperture photometry with simulated data from Figure~\ref{fig:galsim_data}.} The scatter plot shows the differences in magnitudes \mbox{$\Delta m \doteq m_{\widehat{x}} - m_{\operatorname{coadd}}$} as a function of $m_{\operatorname{coadd}}$, where $m_{\operatorname{coadd}}$ and $m_{\widehat{x}}$ are the calibrated magnitudes of corresponding sources from the coadd of the simulated exposures, and from \emph{ImageMM}'s latent image $\widehat{x}$ respectively. Similar to Figure~\ref{fig:photometric_tests_sep}, calibrated magnitudes are consistent up to a value of roughly $25$, beyond which detected sources in $\widehat{x}$ appear brighter than in the coadd.}
    \label{fig:photometric_tests_sep_galsim}
\end{figure*}

Nevertheless, one can still use these latent images in the context of photometric studies, despite their negligible noise levels. We demonstrate this in Figure~\ref{fig:photometric_tests_sep_galsim}, where we performed basic aperture photometry with the simulated imaging data, using exactly the same procedure as described in Section~\ref{ssec:photometric_results_hsc}. 

The photometric analysis confirms that \emph{ImageMM} generates latent images that preserve the fluxes of bright sources, and in which the brightness of faint objects are enhanced relative to their counterparts in traditional coadds. Furthermore, we compared $m_{\widehat{x}}$ (i.e., the magnitudes of sources detected in the latent image $\widehat{x}$ obtained using the simulated images) with the \emph{true} magnitudes used to generate these sources in the simulated images, which we will denote as $m_{\operatorname{true}}$. Using multiple different sets of simulated exposures, we found that the correlation coefficient between $m_{\widehat{x}}$ and $m_{\operatorname{true}}$ was $0.9997$ on average, thereby showing that \emph{ImageMM} produces physically meaningful restorations where photometric measurements agree strongly with the ground truth. These promising results obtained with the simulated data suggest that \emph{ImageMM} holds substantial potential for successful applications with real observational data.

\bibliography{imagemm}{}
\bibliographystyle{aasjournal}

\end{document}